\documentclass[a4paper,twocolumn,11pt,accepted=2022-03-08]{quantumarticle}
\pdfoutput=1
\usepackage{amssymb,amsmath,amsfonts,mathrsfs,bm, bbm, graphicx, epsfig, epstopdf}
\usepackage[version=3]{mhchem}
\usepackage{braket}
\usepackage{hyperref}
\usepackage{color}
\usepackage{csquotes}
\usepackage{float}
\usepackage{subfigure}
\usepackage{soul}

\usepackage[numbers,sort&compress]{natbib}

\begin{document}

\title{Proposal for room-temperature quantum repeaters with nitrogen-vacancy centers and optomechanics}

\author{Jia-Wei Ji}
\email{quantum.jiawei.ji@gmail.com}
\affiliation{Institute for Quantum Science and Technology, and Department of Physics \& Astronomy, University of Calgary, 2500 University Drive NW, Calgary, Alberta T2N 1N4, Canada}

\author{Yu-Feng Wu}
\affiliation{Institute for Quantum Science and Technology, and Department of Physics \& Astronomy, University of Calgary, 2500 University Drive NW, Calgary, Alberta T2N 1N4, Canada}

\author{Stephen C. Wein}
\affiliation{Institute for Quantum Science and Technology, and Department of Physics \& Astronomy, University of Calgary, 2500 University Drive NW, Calgary, Alberta T2N 1N4, Canada}

\author{Faezeh Kimiaee Asadi}
\affiliation{Institute for Quantum Science and Technology, and Department of Physics \& Astronomy, University of Calgary, 2500 University Drive NW, Calgary, Alberta T2N 1N4, Canada}

\author{Roohollah Ghobadi}
\affiliation{Institute for Quantum Science and Technology, and Department of Physics \& Astronomy, University of Calgary, 2500 University Drive NW, Calgary, Alberta T2N 1N4, Canada}

\author{Christoph Simon}
\email{christoph.simon@gmail.com}
\affiliation{Institute for Quantum Science and Technology, and Department of Physics \& Astronomy, University of Calgary, 2500 University Drive NW, Calgary, Alberta T2N 1N4, Canada}

\begin{abstract}
We propose a quantum repeater architecture that can operate under ambient conditions. Our proposal builds on recent progress towards non-cryogenic spin-photon interfaces based on nitrogen-vacancy centers, which have excellent spin coherence times even at room temperature, and optomechanics, which allows to avoid phonon-related decoherence and also allows the emitted photons to be in the telecom band. We apply the photon number decomposition method to quantify the fidelity and the efficiency of entanglement established between two remote electron spins. We describe how the entanglement can be stored in nuclear spins and extended to long distances via quasi-deterministic entanglement swapping operations involving the electron and nuclear spins. We furthermore propose schemes to achieve high-fidelity readout of the spin states at room temperature using the spin-optomechanics interface. Our work shows that long-distance quantum networks made of solid-state components that operate at room temperature are within reach of current technological capabilities. 
\end{abstract}

\maketitle

\section{Introduction}\label{sec:introduction}
The successful implementation of global quantum networks would have many applications such as secure communication~\cite{RevModPhys.74.145}, blind quantum computing~\cite{Barz303}, private database queries~\cite{PhysRevA.83.022301}, ultimately leading to a ``quantum internet''~\cite{Kimble2008,Simon2017,Wehnereaam9288} of networked quantum computers and other quantum devices. This will require photons for establishing long-distance connections, as well as stationary qubits for storing and processing the quantum information. In particular, since quantum information cannot be amplified, quantum repeaters are likely to be required~\cite{RevModPhys.83.33,Muralidharan2016,Simon2017}. Most current approaches to such quantum networks require either vacuum equipment and optical trapping or cryogenic cooling~\cite{Duan2001,Kumar_2019,KimiaeeAsadi2018quantumrepeaters,PhysRevLett.123.063601,Humphreys2018,RevModPhys.83.33,Delteil2016,PhysRevLett.119.010503,Asadi_2020}, which adds significantly to the difficulty of scaling up such architectures. There is notable recent work towards quantum networks with room-temperature atomic ensembles 
~\cite{Borregaard2016,katz2018light,pang2020hybrid,li2021heralding,dideriksen2021room}, but it is also of interest to investigate solid-state approaches, which might ultimately be the most advantageous in terms of scalability.

Nitrogen-vacancy (NV) centers have millisecond-long electron spin coherence times even at room temperature~\cite{Maurer1283,Balasubramanian2009,Bar-Gill2013,PhysRevLett.101.047601}, making them excellent candidates for being stationary qubits in quantum networks~\cite{PhysRevLett.123.063601,Humphreys2018,Hensen2015}. So far, NV-based room-temperature quantum information processors were proposed based on the spin-chain model where the interactions between electron spin qubits are mediated by the nuclear spin chain~\cite{Yao2012} or based on the strongly interacting fluorine nuclear spins~\cite{Cai2013}. It is intriguing to ask whether photonic links can be implemented for NV centers at room temperature. Unfortunately, the phonon-induced broadening of optical transition poses a serious challenge to using NV centers in generating spin-photon entanglement at room temperature~\cite{PhysRevLett.103.256404}. An alternative approach to overcome this problem could be using quantum optomechanics \cite{stannigel2010optomechanical}, where the effective spin-photon coupling is mediated by an ultra-low loss mechanical resonator \cite{Ghadimi764,PhysRevLett.116.147202} to bypass the direct spin-photon interface. It was shown theoretically that this approach allows the emission of highly indistinguishable photons~\cite{Ghobadi2019} at room temperature, which suggests that high-fidelity entanglement creation should be possible as well. Further, this interface allows the freedom of choosing the wavelength of emitted photons. Thus, one could have the emission at telecom band, which is ideal for connecting distant NV centers through optical fibers.  

Nuclear spins in diamond have even longer coherence time at room temperature than the electron spins, exceeding a second~\cite{Maurer1283}. Therefore, these nuclear spins can be used as quantum memory to store the entanglement both at ambient conditions~\cite{Dolde2013}, similar to what is being done at cryogenic temperatures~\cite{PhysRevX.6.021040}. Electron and nuclear spin qubits can be coupled via hyperfine interactions~\cite{Cappellaro2009,Yao2012,Maurer1283}.

\begin{figure*}
\centering
\includegraphics[scale=0.14]{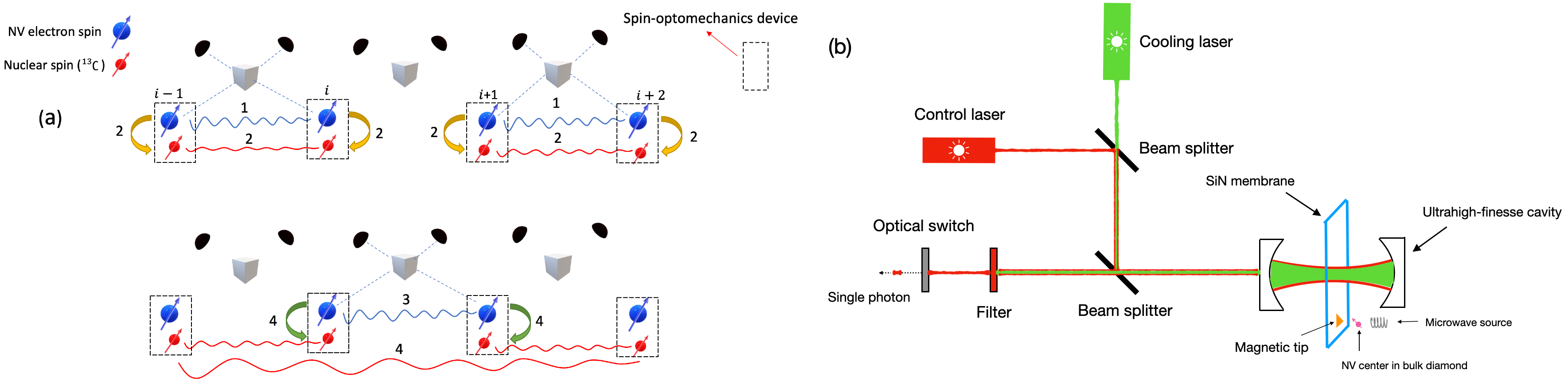}
\caption{(a) Room-temperature quantum repeater architecture. Here, we just show four nodes and three links to demonstrate the basic logic of the quantum repeater protocol, which proceeds in four steps. Step 1 is to generate the entanglement between two remote NV electron spins using the spin-optomechanics interface. Step 2 is the memory mapping, which stores the entanglement between two electron spins into the entanglement between two nuclear spins. Step 3 is the same as step 1 for generating the entanglement between two remote NV electron spins. Step 4 is to perform the entanglement swapping that establishes the entanglement only between the first and the last nuclear spins. (b) Schematic of the spin-optomechanics interface with membrane-in-the-middle design. The optomechanical system consists of a SiN membrane oscillator placed inside the high-finesse cavity. A magnetic tip is attached to this membrane. An NV center in bulk diamond is placed near the tip, such that the oscillator is coupled to the dressed ground states of the NV center. A single telecom photon is produced via the mechanically mediated interaction between the control laser and the dressed NV center, while the cooling laser is on to keep the membrane oscillator near its ground state.}
\label{Figure 1}
\end{figure*}
 
Based on the above line of thought, we here propose a room-temperature quantum repeater architecture based on NV centers and optomechanics.  In our proposal the entanglement between two distant NV electron spins is established via photons following the Barrett-Kok scheme~\cite{PhysRevA.71.060310,Bernien2013,Hensen2015}. We apply the photon number decomposition method~\cite{PhysRevA.102.033701} to quantify and analyze the entanglement generation efficiency and fidelity. Mapping of the electron spin entanglement onto nuclear spins is achieved via performing CNOT gates and electron spin readout through the spin-optomechanics interface. Finally, entanglement swapping is done using the same gate operations assisted by the readout of electron spin and nuclear spin states. The quasi-deterministic gate operations allow us to distribute the entanglement in the nesting-level free manner which outperforms the rates of other conventional nested repeater protocols. Moreover, multiplexing is an indispensable part of our proposal, which allows for feasible fidelities of entanglement distributed at long distances.

This paper is organized as follows. In Sec.~\hyperref[sec:quantum repeater]{2}, we introduce the quantum repeater architecture, including the spin-optomechanics interface, as well as entanglement generation, entanglement storage in nuclear spins, and entanglement swapping. The NV electron spin readout at room temperature is discussed in Sec.~\hyperref[sec: readout]{3}. Sec.~\hyperref[sec:rates]{4} discusses the repeater rates and fidelities. Sec.~\hyperref[sec:implementation]{5} gives more details on implementation. We conclude and provide an outlook in Sec.~\hyperref[sec:conclusion]{6}.

\section{Quantum repeater architecture} 
\label{sec:quantum repeater}
The diagram in Fig.~\ref{Figure 1}(a) illustrates the basic steps and components for building a room-temperature quantum repeater architecture based on spin-optomechanics systems. A typical quantum repeater features two basic ingredients: the entanglement generation between two remote memories, and the entanglement swapping between two local memories to propagate it further~\cite{Kimble2008,Simon2017}. Here, our physical systems also have these two components, and they can operate at room temperature. One crucial component of our proposal is the spin-optomechanics interface which was first proposed by R. Ghobadi et.al.~\cite{Ghobadi2019}. Moreover, our proposal features two kinds of qubits: the NV electron spins serve as communication qubits, and the nuclear spins serve as memory qubits for storing the entanglement because they have long coherence time even at room temperature~\cite{Maurer1283,Dolde2013}. At cryogenic temperature, experimental realizations of such diamond-based nuclear-spin memories have already been demonstrated~\cite{PhysRevX.6.021040,Dolde2013}.

This section is dedicated to the basic structure and components of our proposed architecture. We start with the introduction to the spin-optomechanics interface~\cite{Ghobadi2019}, and then quantify the efficiency and fidelity of entanglement generation between two remote nodes based on the recently developed photon number decomposition method~\cite{PhysRevA.102.033701}. Then we discuss entanglement storage and swapping under ambient conditions. The application of the spin-optomechanics interface for the electron spin state readout at room temperature, which serves as a crucial ingredient in the proposed architecture, is discussed in the next section.

\subsection{Spin-optomechanics interface}\label{subsec: IIA}
The schematic of spin-optomechanics interface is shown in Fig.~\ref{Figure 1}(b). There are three main components in the system: the NV electron spin, the mechanical oscillator (SiN membrane) and the high-finesse optical cavity. The NV electron spin is coupled to the mechanical oscillator via a magnetic tip that is attached to the oscillator, which requires the magnetic field gradient to produce the strong spin-mechanics coupling rate $\lambda$~\cite{Ghobadi2019}. The red-detuned control laser is used to induce the optomechanical coupling rate $g$.  The NV electron spin must be tuned to be resonant with the red-detuned control laser so that a single spin-excitation would be converted a single photon emitted at the cavity frequency via the mechanical oscillator. However, when the control laser is red-detuned from the cavity, it also starts to cool the mechanical oscillator via the phonon sideband. This converts phonons to single photons at the cavity frequency as well, which causes a thermal noise that degrades the quality of the single photon from the NV electron spin. In order to reduce this noise, we detune the control laser far from the phonon sideband $\omega_m$. Since the control laser is detuned far from the phonon sideband, it is ineffective at cooling the mechanical oscillator. Hence, we introduce a different laser on resonance with the mechanical oscillator to efficiently cool it \cite{Ghobadi2019}.

\begin{figure}
\centering 
\includegraphics[scale=0.32]{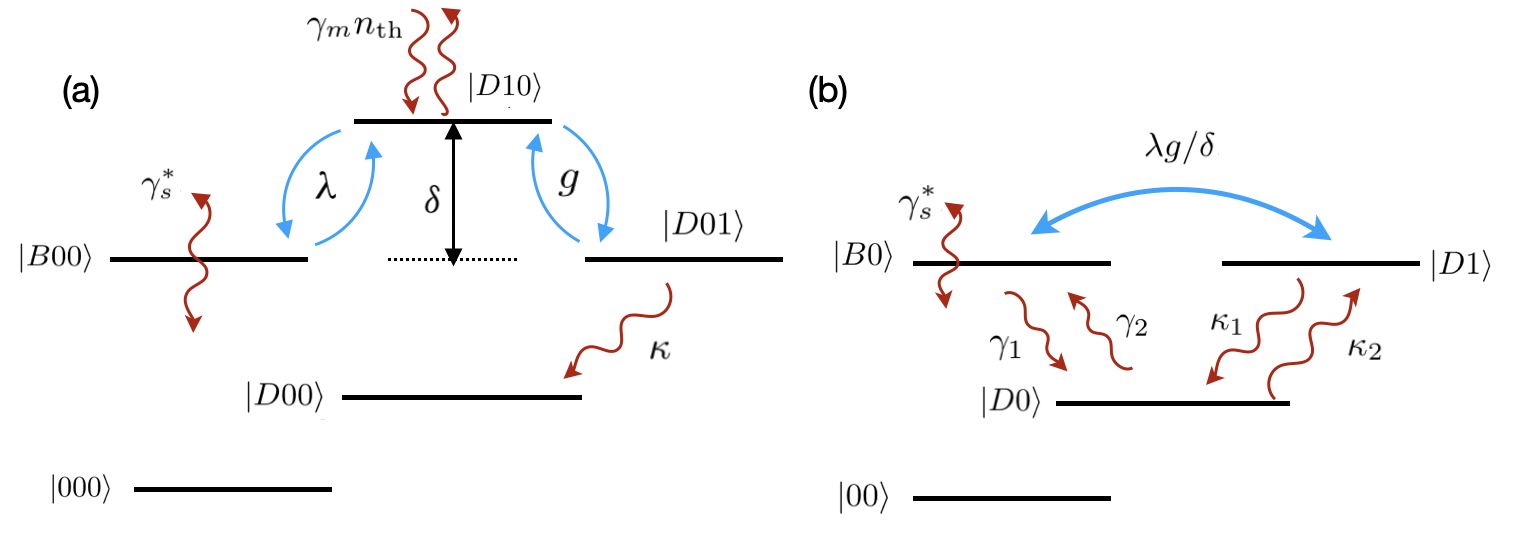}
 \caption{(a) The level diagram illustrates the coupling between the excited dressed NV electron spin state and the mechanical phonon with the rate $\lambda$, and the coupling between the mechanical phonon and the cavity photon with rate $g$. Coupled states are denoted as $\ket{\text{spin},\text{mechanics},\text{cavity}}$. A single photon is generated via the indirect coupling between the spin and cavity mode through the oscillator, and is then released by the cavity at the rate $\kappa$, leaving the whole system in $\ket{D00}$. The dressed spin state has dephasing rate $\gamma^*_s$, and the mechanical oscillator is dissipatively driven by the environment with the rate $\gamma_m n_{\text{th}}$. (b) The schematic of four-level spin-cavity system after the adiabatic elimination of oscillator mode. The effective coupling strength between the cavity and the NV spin is $\lambda g/\delta$. This effective spin-cavity system has five effective decoherence rates: the pure spin dephasing rate $\gamma^*_{\text{s}}$, the mechanically-induced thermal decay and excitation rates $\gamma_1$ and $\gamma_2$ for the spin, and the effective decay rate $\kappa_1$ and mechanically-induced thermal excitation rate $\kappa_2$ for the cavity mode.}
 \label{Figure 2}
 \end{figure}

The triplet NV electron spin state $\{\ket{0},\ket{-1},\ket{+1}\}$ is under the dressing of a microwave source~\cite{Ghobadi2019}, which form a three-level dressed spin states $\{\ket{0},\ket{D},\ket{B}\}$ that are noise-protected from the nuclear-spin bath~\cite{PhysRevLett.114.120501}. Only the bright state $\ket{B}=(\ket{+1}+\ket{-1})/\sqrt{2}$ and the dark state $\ket{D}=(\ket{+1}-\ket{-1})/\sqrt{2}$ couple to the mechanical oscillator with the rate $\lambda$. The states $\ket{+1}$ and $\ket{-1}$ are two of the triplet ground states of the NV center. The transition frequency between $\ket{B}$ and $\ket{D}$ is $\omega_q$, which is tuned to be the same as the control laser via controlling the Rabi frequency of the microwave dressing source. The detuning $\delta$ between the red-detuned control laser $\omega_q$ and the phonon sideband $\omega_m$ is $\delta=\omega_m-\omega_q$. The level diagram of this spin-optomechanics system is shown in Fig.~\ref{Figure 2}(a).

Then, the system Hamiltonian is given by ($\hbar=1$)
\begin{equation}
\hat{H}=\omega_{\textrm{q}}(\hat{\sigma}_+\hat{\sigma}_-+\hat{a}^{\dagger}\hat{a})+\omega_{\textrm{m}}(\hat{b}^{\dagger}\hat{b}+\hat{c}^{\dagger}\hat{c})+\hat{H}_{\textrm{I}},
\label{eq:somHam}
\end{equation}
where $\hat{\sigma}_{-}=\ket{D}\bra{B}$ is the lowering operator for the dressed NV spin states, and $\hat{a}$ and $\hat{c}$ are the control cavity mode and cooling cavity mode respectively, and $\hat{b}$ is the oscillator mode. $\hat{H}_{\textrm{I}}$ stands for the interaction term, and it takes the following form: 
\begin{equation}
\hat{H}_{\textrm{I}}=(\lambda\hat{\sigma}_-+g\hat{a}+g_{\textrm{c}}\hat{c})(\hat{b}^{\dagger}+\hat{b})+\textrm{H.c.},
\label{eq:somint}
\end{equation}
where $\lambda$ is the spin-mechanics coupling strength, $g$ is the control optomechanical coupling rate, and $g_{\textrm{c}}$ is the cooling optomechanical coupling rate.    

Under the condition that $\delta\gg\{\lambda,g\}$, and the cooling mode significantly reduces the thermal noise from the mechanical oscillator, making it near the ground state~\cite{Ghobadi2019}, it is valid to adiabatically eliminate the $\delta$-detuned mechanical phonon mode to achieve the effective coupling between the dressed spin state and a cavity photon~\cite{Ghobadi2019,PhysRevApplied.4.044003}. The cooling mode can also be ignored as it cools the mechanical oscillator, converting phonons to photons that are emitted at a different frequency than the desired single photon from the NV spin. The effective coupling rate is $\lambda g/\delta$ as indicated by the blue arrow in Fig.~\ref{Figure 2}(b). After adiabatic elimination and rotating-wave approximation ($\delta\ll\omega_q,\omega_m$), the simplified Hamiltonian is given by~\cite{Ghobadi2019,NoteX}
\begin{equation}
\hat{H}_{\text{eff}}=\frac{g^2}{\delta}\hat{a}^{\dagger}\hat{a}+\frac{\lambda^2}{\delta}\hat{\sigma}_+\hat{\sigma}_-+\Omega(\hat{a}^{\dagger}\hat{\sigma}_-+\hat{a}\hat{\sigma}_+),
\label{eq:adeha}
\end{equation}
where $\Omega=\lambda g/\delta$ is the effective coupling strength between the cavity photon and NV bright state. Although this system is a three-level system containing two coupled ground states of NV spin $\{\ket{D},\ket{B}\}$ and the cavity mode, it is convenient to include the uncoupled ground state $\ket{0}$ in the system for the later analysis. From now, we call this system a four-level system. Then, the corresponding effective master equation is given by~\cite{NoteX}
\begin{equation}
\begin{aligned}
\dot{\hat{\rho}}=&-i[\hat{H}_{\text{eff}},\hat{\rho}]+\kappa_1\mathcal{D}[\hat{a}]\hat{\rho}+\gamma^*_s\mathcal{D}[\hat{\sigma}_z]\hat{\rho}\\
&+\gamma_1\mathcal{D}[\hat{\sigma}_-]\hat{\rho}+\gamma_2\mathcal{D}[\hat{\sigma}_+]\hat{\rho}+\kappa_2\mathcal{D}[\hat{a}^{\dagger}]\hat{\rho}, 
\end{aligned}
\label{eq:ademaster}
\end{equation}
where $\kappa_1=\kappa+g^2\gamma_m(n_{\textrm{th}}+1) /\delta^2$ is the effective cavity decay rate with original cavity decay rate $\kappa$, and $\kappa_2=g^2n_{\textrm{th}}\gamma_m/\delta^2$ is the mechanically-induced thermal excitation rate for the cavity photon with the oscillator damping rate $\gamma_m$ and the average phonon number $n_{\textrm{th}}$ determined by the environment temperature, and $\gamma^*_s$ is the pure spin dephasing rate, and $\gamma_1=\lambda^2\gamma_m(n_{\textrm{th}}+1)/\delta^2$, $\gamma_2=\lambda^2n_{\textrm{th}}\gamma_m/\delta^2$ are the mechanically-induced thermal decay and excitation rates for the NV spin state, respectively. Here $\mathcal{D}[\hat{A}]\hat{\rho}=\hat{A}\hat{\rho}\hat{A}^\dagger-\hat{A}^\dagger\hat{A}\hat{\rho}/2-\hat{\rho}\hat{A}^\dagger\hat{A}/2$. The inherent NV spin flip-flop rate is ignored because it is much smaller than the pure spin dephasing rate $\gamma^*_s$ even at ambient temperature~\cite{Balasubramanian2009}.

\subsection{Entanglement generation}

Step 1 in Fig.~\ref{Figure 1} is to generate entanglement between two remote NV electron spins at room temperature. This can be achieved using the protocol described in Sec.~\ref{subsec: IIA}. Photons with high indistinguishability, brightness and purity can be produced using this spin-optomechanics interface at room temperature~\cite{Ghobadi2019}. Each of the two spin-optomechanical interfaces can be modeled as described in the previous section.

If the initial state of the NV center is prepared as $(\ket{B}+\ket{0})/\sqrt{2}$, a single photon would be released from $\ket{B0}$ at the cavity frequency via the effective coupling between $\ket{B0}$ and $\ket{D1}$. Therefore, a spin-photon entangled state $(\ket{D1}+\ket{00})/\sqrt{2}$ is created. Then, after interfering the photonic modes from each interface at a beam splitter, detection of a single photon projects the two spins into an entangled state. Here, we propose to use the spin-time bin protocol (the Barrett-Kok scheme) to generate the entanglement between two distant nodes, which is much more robust against some important errors such as photon loss, detector loss and cavity parameters mismatch compared the single-photon detection scheme~\cite{Bernien2013,PhysRevA.71.060310}. In this protocol, two rounds of single-photon detection are required. After the first round, we flip the spin states $\ket{D}$, $\ket{0}$ of both systems and re-excite $\ket{D}$ to $\ket{B}$. The detection of two consecutive single photons (one at each round), will then project the joint state of the quantum systems onto a Bell state. Depending on which detectors click in these two rounds, we obtain two Bell states $\ket{\psi_\pm}=(\ket{D0}\pm\ket{0D})/\sqrt{2}$ with a 50$\%$ total probability.

\begin{figure}[t]
\centering
\includegraphics[scale=0.7]{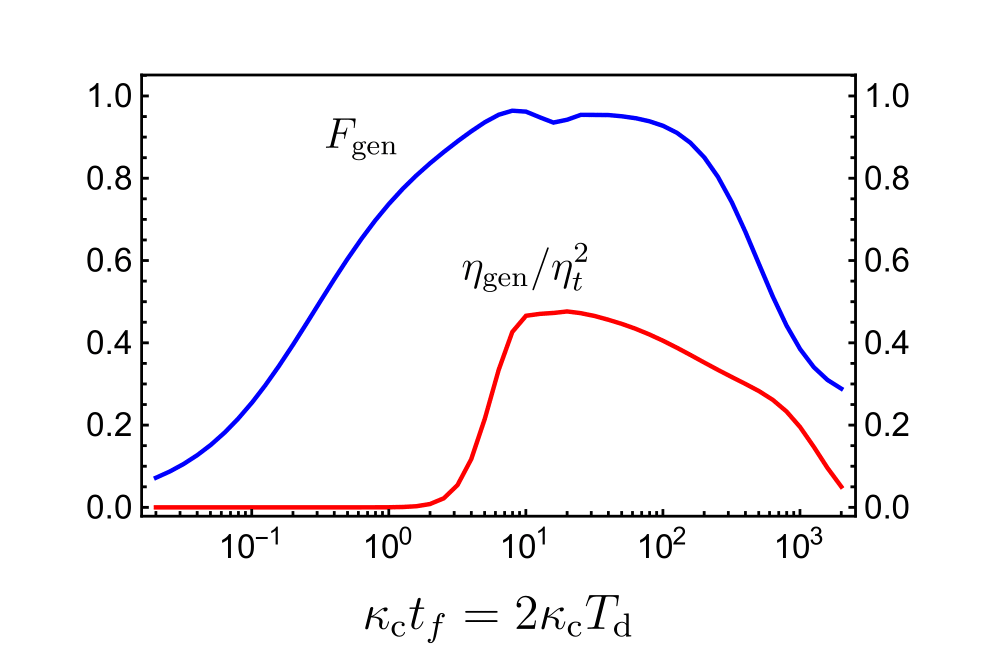}
\caption{Entanglement generation fidelity $F_{\text{gen}}$ and efficiency $\eta_{\text{gen}}/\eta^2_t$ for a single link as a function of protocol time $t_\text{f}$. The mechanically-induced initial thermal noise in the cavity is modeled as dynamical dark counts as described in the text, while the detector dark count rate is set to 10 Hz~\cite{Bai2017}. The detection time window for each time bin $T_\text{d}$ is set to be equal to half the total detection time window: $t_\text{f}=2T_\text{d}$. Due to the loss in the channel, it is difficult to see the efficiency curve so it is divided by the factor $\eta^2_t=\text{exp}(-L_0/L_{\text{att}})$, where $L_0=100$ km is the length of the link, and $L_{\text{att}}=22$ km is fiber attenuation distance of telecom photons. The peak value of the fidelity curve $F_{\text{gen}}$ is around 97\%. All parameters are chosen to be the same for both spin-optomechanics systems and similar to those in Ref.~\cite{Ghobadi2019}, where the parameters are optimized for achieving high indistinguishability and single-photon purity: $\lambda=g=2\pi\times 100$ kHz, $\delta=2\pi\times 1$ MHz, $Q_m=3\times10^9$, $\kappa_1=2\Omega=2\pi\times20~\text{kHz}$, $\gamma_s^\star=0.01\kappa_1$~\cite{Balasubramanian2009}, and $\gamma_{\text{1}}=\gamma_{\text{2}}=1.0\times10^{-3}\kappa_1$.}  
\label{Figure 4}
\end{figure}

Due to the existing mechanically-induced cavity emission at room temperature, the initial state of the cavity is not perfectly the vacuum state. A more precise initial state can be obtained by solving the steady state of cavity mode with only the optomechanical coupling $g$ turned on. Then, the initial state of the cavity is given by \cite{NoteX}
\begin{equation}
\rho_{ic}=\frac{\kappa_1-2\kappa_2}{\kappa_1-\kappa_2}\ket{0}\bra{0}+\frac{\kappa_2}{\kappa_1-\kappa_2}\ket{1}\bra{1},    
\end{equation}
where $\kappa_1\gg\kappa_2$. The mechanically-induced initial thermal occupation $\kappa_2/(\kappa_1-\kappa_2)$ is quite small, which is estimated to be around 0.1\% using the parameters in Fig.~\ref{Figure 4}. Since this thermal occupation is so small, and it does not affect the quantum system dynamics significantly, we can treat its contributions classically by modelling it as dark counts to simplify the calculations~\cite{PhysRevA.102.033701}. This dark count rate is given by $\mathcal{D}_{\text{th}}=\kappa_1\kappa_2/(\kappa_1-\kappa_2)$. Therefore, we start with the initial state of the system: $\hat{\rho}(t_0)=\ket{\psi(t_0)}\!\bra{\psi(t_0)}$ where $\ket{\psi(t_0)}=(1/2)(\ket{0,0}+\ket{B,0})^{\otimes 2}$. Under this approximation, the mechanically-induced thermal excitation rate in the cavity mode can be set to 0 in Eq.~(\ref{eq:ademaster}), i.e., $\kappa_2=0$. In this way, the total number of quantum states to simulate is reduced. 

Now, in order to quantify the entanglement fidelity and efficiency, we follow the photon number decomposition method developed in~\cite{PhysRevA.102.033701} to compute the time dynamics. The basic idea of this method is to decompose the master equation dynamics into evolution conditioned on single photon detection, which can be done by rewriting the master equation of the whole system (in this case two distant spin-optomechanical systems) as follows:
\begin{equation}
\label{meqC}
    \dot{\rho}=\mathcal{L}_0\hat{\rho}(t) +\sum_{i=1}^2\mathcal{S}_{i}\hat{\rho}(t),
\end{equation}   
where $\mathcal{L}_0=\mathcal{L}-\sum_{i=1}^2\mathcal{S}_i$ with $\mathcal{L}$ being the Liouville superoperator that contains all the dynamics of this composite system, and $\mathcal{S}_i\hat{\rho}=\hat{d}_i\hat{\rho}\hat{d}_i^\dagger$ is the collapse superoperator of the source field $\hat{d}_i$ at the $i^\text{th}$ single-photon detector \cite{PhysRevA.102.033701}. As can be seen, at a given detection time window $t_f$ if there is no photon detected, then the system evolves only subject to $\mathcal{L}_0$, but if there is a photon detected during this time window, then we apply the collapse superoperator to the system. Moreover, as the final state of the system depends on the detected photon count, we would obtain a set of different states, which we call conditional states.

In the Barrett-Kok scheme, there are four possible detected photon counts: $\{\mathbf{n}_l,\mathbf{n}_\text{e}\}=\{(1,0),(1,0)\}$, $\{(1,0),(0,1)\}$, $\{(0,1),(1,0)\}$, and $\{(0,1),(0,1)\}$ where $\mathbf{n}_l$ and $\mathbf{n}_e$ stand for the photon count in the early and late detection time window, and each can take two possible outcomes $(1,0),(0,1)$ which correspond to the click in the left detector and the right detector as shown in Fig. \ref{Figure 1}. Thus, the entanglement generation efficiency and the entanglement generation fidelity can be defined in the following way: 
\begin{equation}
\begin{aligned}
\eta_{\text{gen}}&=\text{Tr}[\hat{\rho}(t_f)]=\sum_{\mathbf{n}}^{}{\text{Tr}[\hat{\rho}_{\mathbf{n}}(t_f)]}\\
F_{\text{gen}}&= \frac{1}{4}\sum_{\mathbf{n}}^{}{\frac{\bra{\psi_\pm}\hat{\rho}_{\mathbf{n}}(t_f)\ket{\psi_\pm}}{\text{Tr}[\hat{\rho}_{\mathbf{n}}(t_f)]}},
\end{aligned}
\end{equation} 
where $\mathbf{n}$ stands for the detected photon count as mentioned above, and we use $\ket{\psi_+}$, when $\mathbf{n}=\{(1,0),(1,0)\}, \{(0,1),(0,1)\}$, otherwise we use $\ket{\psi_-}$. Further, due to dark counts (both from detectors and the initial thermal occupation as mentioned above), zero or single-photon conditioned states would give spurious photon counts. This imperfection is also taken into account when estimating the the entanglement generation fidelity and efficiency, which is discussed in more detail in \cite{PhysRevA.102.033701}.

Fig.~\ref{Figure 4} shows the entanglement generation fidelity $F_{\text{gen}}$ and efficiency curves $\eta_{\text{gen}}/\eta^2_t$ for the effective spin-cavity system described by Eq.~(\ref{eq:ademaster}) over the total detection time window $t_{\text{f}}$ for a link of 100 km. $T_\text{d}$ is the detection time window for each time bin, which is set to be half the total detection time window $t_{\text{f}}$. The loss in the channel degrades the entanglement efficiency in proportion to the square of the transmission rate, i.e., $\eta^2_t=\text{exp}(-L_0/L_{\text{att}})$, which makes the efficiency curve difficult to see, so it is divided by this factor. We assume a dark count rate of 10 Hz, which is predicted to be achievable for photons in the telecom band using up-conversion single photon detectors (USPDs) in the free-running regime 
~\cite{Bai2017} (which do not require cryogenic cooling).  After taking the loss in the channel into account, this detector dark count rate is comparable to the rate $\mathcal{D}_{\text{th}}\sim 100$ Hz. This type of detector is also predicted to have low afterpulsing probability~\cite{Bai2017}, making afterpulsing negligible in estimating entanglement fidelity and efficiency. For the detection efficiency, we consider 45$\%$ \cite{Bai2017}, which is later used in the readout fidelity estimates and the repeater rates calculations.

Fig.~\ref{Figure 4} shows that the efficiency degrades gradually after it reaches the maximum due to the thermal-induced flip-flop effect between the bright and dark states. Under the influence of flip-flop effect, both systems continue to emit photons, resulting in the probability of detecting only two photons to vanish when the detection time $t_{\text{f}}$ goes to infinity. Likewise, the fidelity decreases after it reaches the maximum, and it starts with fairly low values due to the small signal-to-noise ratio in the beginning. If we choose to terminate the measurement at a proper time as $\kappa_1t_{\text{f}}\sim 10$, then the fidelity is approaching 97\% at room temperature.

One can obtain approximate analytical expressions for the entanglement fidelity and efficiency by following the methods developed in ~\cite{PhysRevLett.114.193601,PhysRevB.97.205418}. In the incoherent regime ($2\Omega\leq\kappa+2\gamma^*_s+2\Gamma_{\text{th}}$), we can model this four-level system as a three-level system with the effective emission rate by adiabatically eliminating the spin-photon coherence~\cite{Ghobadi2019,NoteX}:
\begin{equation}
R=\frac{4\Omega^2}{(\kappa+2\gamma^\star_s+2\Gamma_{\text{th}})},    
\end{equation}
where $\Gamma_{\text{th}}=\lambda gn_{\text{th}}\gamma_m/\delta^2$ is the thermal-induced noise. By applying the photon number decomposition method to this spin-optomechanics system~\cite{PhysRevA.102.033701}, we get the entanglement generation efficiency in the Barrett-Kok scheme:
\begin{equation}
\eta_{\text{BK}}=\frac{\eta^2_t}{2}(1-e^{\frac{-t_\text{f}}{2}R})^2,
\end{equation}
where $R$ is the effective emission rate for each system, and $\eta_t$ is the transmission rate in the channel. This is proportional to the product of the two total emission intensities from the two emitters. However, for the room-temperature case where the cavity starts with a small thermal occupation, a more precise expression of the efficiency is given by taking the dark counts into consideration as discussed in~\cite{PhysRevA.102.033701}.

The entanglement generation fidelity $F_{\text{BK}}$ is then given by~\cite{PhysRevA.102.033701}
\begin{equation}
F_{\text{BK}}=\frac{1}{2}\left(1 + \frac{1}{2\eta_\text{BK}}|\tilde{C}(t_{\text{f}})|^2\right),    
\end{equation} 
where $\tilde{C}(t_{\text{f}})$ takes the following form
\begin{equation}
\tilde{C}(t_{\text{f}}) =\frac{\eta_t R}{R_{\text{tot}}}\left(1-e^{-\frac{1}{2}t_{\text{f}}R_{\text{tot}}}\right),\end{equation}
where $R_{\text{tot}}=R+2\gamma^\star_{s}$ is the spectral width of the emitted photons for both systems.~This fidelity equation is the upper bound for the cryogenic temperature case when there is only optical dephasing. For the room-temperature case, one needs to take into account the mechanically-induced thermal contribution in the cavity and the mechanically-induced spin flip-flop effect, which makes the precise analytical fidelity expression very difficult to obtain.

\subsection{Entanglement mapping}\label{subsec: IIB}

After the successful entanglement generation, we need to store the entanglement between two remote NV electron spins in nuclear spins via performing memory swapping between an electron spin and a nuclear spin at both ends of the link as indicated by two yellow arrows in Fig.~\ref{Figure 1}. This operation is achieved through performing a $\mbox{C}_n\mbox{NOT}_e$ gate between the electron and nuclear spins plus the measurement of the state of the electron spin.

\begin{figure}
\centering 
\includegraphics[scale=0.5]{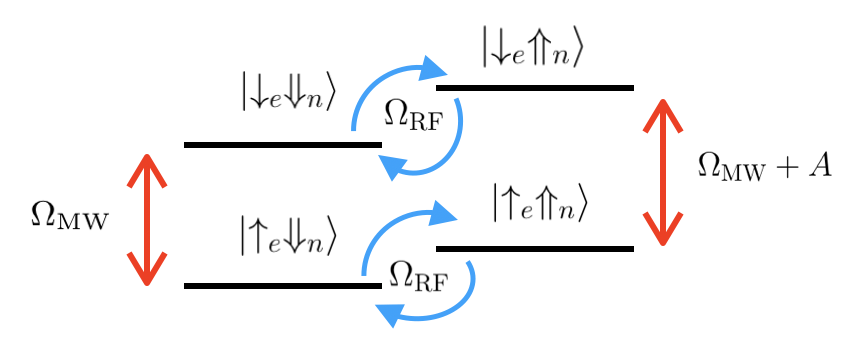}
 \caption{The NV center with a \ce{^{13}_{}C} can be modeled as a four-level system. Nuclear spin sublevels $\ket{\Uparrow_n}$ and $\ket{\Downarrow_n}$ are addressed by RF radiation with Rabi frequency $\Omega_{RF}$. The electronic spin sublevels are driven via a microwave field $\Omega_{\text{MW}}$ but when the electron spin is $\ket{\downarrow_e}$, the microwave field has relative detuning given by hyperfine interaction $A$.}
 \label{Figure 5}
 \end{figure}

Assuming that $\ket{\psi^+}$ is obtained in step 1, since quantum systems are in the dressed basis $\{\ket{B},\ket{D},\ket{0}\}$, we need to bring them back to the original basis $\{\ket{+1},\ket{-1},\ket{0}\}$ by turning off the microwave source adiabatically. Then, $\ket{D}$ returns to $\ket{-1}$ and $\ket{0}$ remains the same. Here, we denote $\{\ket{-1},\ket{0}\}$ as $\{\ket{\uparrow_e},\ket{\downarrow_e}\}$ for the electron spin. Then, we prepare the nuclear spin in the superposition of the spin-up and spin-down states by applying a $\pi/2$ RF pulse to the nuclear spin that is initially polarized to the spin-down state via the combination of optical, microwave, and RF fields as discussed in \cite{Jiang267}. There are several options for nuclear spins in diamond such as \ce{^{14}_{}N}~\cite{Wrachtrup2010} and \ce{^{15}_{}N}~\cite{doi:10.1063/1.4748280}. Here, we use \ce{^{13}_{}C} as the nuclear spin in an isotopically purified sample, which has the nuclear spin $I=1/2$~\cite{Balasubramanian2009,Maurer1283,Kalb928}. The state is then given by
\begin{equation}
\begin{aligned}
\frac{1}{\sqrt{2}}(\ket{\Downarrow_n}+\ket{\Uparrow_n})&\otimes      \frac{1}{\sqrt{2}}(\ket{\downarrow_e\uparrow_e}+\ket{\uparrow_e\downarrow_e})\\
&\otimes\frac{1}{\sqrt{2}}(\ket{\Downarrow_n}+\ket{\Uparrow_n}),
\end{aligned}
\end{equation}
where $\ket{\Downarrow_n}$ and $\ket{\Uparrow_n}$ correspond to $m_{I}=-1/2$ and $m_{I}=+1/2$ individually. Now, a $\mbox{C}_n\mbox{NOT}_e$ gate can be performed between the electron and nuclear spins using the hyperfine interaction between them. Fig.~\ref{Figure 5} shows the hyperfine structure for performing two-qubit gates between the electron spin and the nuclear spin and one-qubit gates on each of them individually.

The electron-nuclear spin Hamiltonian is given by
\begin{equation}
H_{e,n}=\Delta_0S^2_z+\mu_eBS_z+\mu_nBI_z+AS_zI_z,
\end{equation}
with the zero-field splitting $\Delta_0$=2.87 GHz, the electronic spin gyromagnetic ratio $\mu_e=-2.8$ MHz/Gauss, the nuclear spin gyromagnetic ratio $\mu_n$ =1.07 kHz/Gauss, the external magnetic field $B$ is applied along the symmetry axis of the NV, and the hyperfine coupling $A$ ranges from tens of kHz to 100 MHz for a \ce{^{13}_{}C} nuclear spin~\cite{Smeltzer_2011,Neumann1326,Maurer1283}. The $\mbox{C}_n\mbox{NOT}_e$ gate can be implemented by a Ramsey sequence on the electron spin at room temperature, where the free precession time is chosen to be $t=\pi/A$ with the magnetic field of several hundred Gauss~\cite{Maurer1283,Wrachtrup2010,Jiang267}.
The efficient realization of the CNOT gate with fidelity of $99.2\%$ at ambient conditions has been demonstrated using composite pulses and an optimized control method~\cite{Rong2015} as well as the dynamical decoupling technique~\cite{DeLange2010,PhysRevLett.105.200402,PhysRevB.83.081201}. The dynamical decoupling technique is also important in the entanglement generation where the electron spin can be decoupled from the nuclear spin bath to have millisecond-long coherence time at room temperature~\cite{Bar-Gill2013,PhysRevB.83.081201}. However, in our entanglement generation step the NV electron spin is in dressed states under a far-detuned microwave source, which itself is already robust against the nuclear-bath-induced noise~\cite{PhysRevLett.114.120501,PhysRevB.79.041302}.

Two $\mbox{C}_n\mbox{NOT}_e$ gates on both ends of the link lead to a four-qubit entangled state. So the projective measurement in the Z basis on the state of the electron spin is required to complete the entanglement storage, which projects this four-qubit entangled state to an entangled state of the nuclear spins. Typically, fluorescence detection can be used to determine the state of the electron spin after the projective measurement at low temperature around 4K with good fidelity~\cite{Blok2014}, which enables the cryogenic-temperature entanglement storage in nuclear spins~\cite{PhysRevX.6.021040,Kalb928}. Unfortunately, at room temperature the intensity of electronic spin-up and spin-down states only differ by roughly a factor of 2 due to the fact that the phonon-induced broadening greatly diminishes the resolution of these two Zeeman states~\cite{Wrachtrup2010}. Thus, the past decade has seen a great deal of experimental efforts put into solving this problem~\cite{Jiang267,Shields2015,PhysRevB.100.125436,Siyushev2019}. In Sec.~\ref{sec: readout}, we propose two electron spin readout schemes based on the spin-optomechanics system.

\subsection{Entanglement swapping}
\label{subsec:entswap}
After mapping the entanglement to the nuclear spins, the electron spins are free and we can use them again to generate entanglement between the electron spins $i$ and $i+1$. This is done in step 3 as illustrated in Fig.~\ref{Figure 1}. Then, the entanglement swapping is achieved as follows: a $\mbox{C}_n\mbox{NOT}_e$ gate at each endpoint of this link is applied, giving us an entangled state of these six spins. Via performing measurements on the electron spin in the Z basis, one ends up obtaining an entangled state of four nuclear spins. Depending on the measurement outcomes, one gets different entangled states. Here, we assume that we get the following four-qubit entangled state: 
\begin{equation}
\frac{1}{\sqrt{2}}(\ket{\Uparrow_n\Uparrow_n\Uparrow_n\Uparrow_n}+\ket{\Downarrow_n\Downarrow_n\Downarrow_n\Downarrow_n}).
\end{equation}

In order to complete the entanglement swapping, i.e. to only entangle nuclear spins $i-1$ and $i+2$, one still needs to disentangle two nuclear spins $i$ and $i+1$ in between. This can be done by measuring them in the $X$ basis but unfortunately, one cannot optically read out the nuclear spin directly. However, it turns out that the nearby electron spins can be used to indirectly read out the nuclear spin state~\cite{Wrachtrup2010,PhysRevLett.110.060502}. The basic idea is as follows: first, a Hadamard gate is performed on the nuclear spins $i$ and $i+1$ individually by applying a $\pi/2$ RF pulse to make $\ket{\Uparrow_n}\rightarrow 1/\sqrt{2}(\ket{\Downarrow_n}+\ket{\Uparrow_n})$ and $\ket{\Downarrow_n}\rightarrow 1/\sqrt{2}(\ket{\Downarrow_n}-\ket{\Uparrow_n})$.

Second, the electron spin nearby is initialized to $\ket{\uparrow_e}$, and we again perform a $\mbox{C}_n\mbox{NOT}_e$ gate, mapping the nuclear spin state to the electron spin state. Therefore, the readout of the nuclear spin could be achieved by performing the measurements in the Z basis on the electron spin, followed by the readout of the measurement outcome which is discussed in detail in Sec.~\ref{sec: readout}. The post-measurement state is given by
\begin{equation}
\begin{cases}
\frac{1}{\sqrt{2}}(\ket{\Uparrow_n\Uparrow_n}-\ket{\Downarrow_n\Downarrow_n}) &\uparrow_e\downarrow_e\mbox{or}\downarrow_e\uparrow_e\\
\frac{1}{\sqrt{2}}(\ket{\Uparrow_n\Uparrow_n}+\ket{\Downarrow_n\Downarrow_n}) &\uparrow_e\uparrow_e\mbox{or}\downarrow_e\downarrow_e,
\end{cases}
\end{equation}
where the final state depends on the outcomes of the electron spins readout. Therefore, nuclear spins $i-1$ and $i+2$ are entangled as indicated by the long red wavy line in Fig.~\ref{Figure 1}(b). As we can see, the entanglement swapping process is in fact equivalent to the entanglement mapping process plus the readout of two nuclear spins. 

\section{The electron spin readout}
\label{sec: readout}
Applying previously proposed readout methods to our system is quite challenging since they require extra techniques and apparatus such as using nuclear spin ancillae, spin-to-charge conversion~\cite{Shields2015} and photoelectrical imaging~\cite{Siyushev2019} to achieve a high-fidelity readout of electron spin at room temperature. Hence, we propose to read out the electron spin state at room temperature using the spin-optomechanics interface. In this section, two intensity-based readout schemes are proposed to distinguish the electron spin state at room temperature.

\subsection{Readout scheme using periodic driving pulses}
\label{subsec: intensity}

In the readout scenario, the aim is to distinguish the states $\ket{0}$ or $\ket{D}$. The intuitive idea is to perform a $\pi$ pulse on the transition between $\ket{B}$ and $\ket{D}$, which will excite the state $\ket{D}$ to $\ket{B}$ while keeping the state $\ket{0}$ unchanged. Then the state $\ket{B}$ will decay back to $\ket{D}$ according to the process described in Fig.~\ref{Figure 2}(a) and will emit a single photon. By measuring a single photon, we can determine that the state is initially in the state $\ket{D}$ or $\ket{0}$. However, measuring a single photon may not be the optimal way to distinguish these two spin states due to the photon loss in the channel and the dark counts in detectors. Therefore, we provide two extended readout schemes, the periodic driving scheme and the continuous driving scheme to achieve the high-fidelity readout of NV electron spin states.

\begin{figure}[t]
\centering
$\hspace{-75mm}\mathbf{(a)}$\\
\includegraphics[scale=0.27]{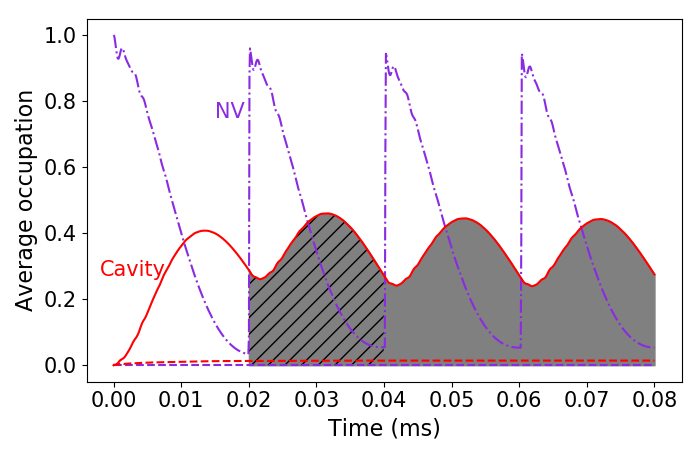}\\
$\hspace{-75mm}\mathbf{(b)}$\\
\includegraphics[scale=0.27]{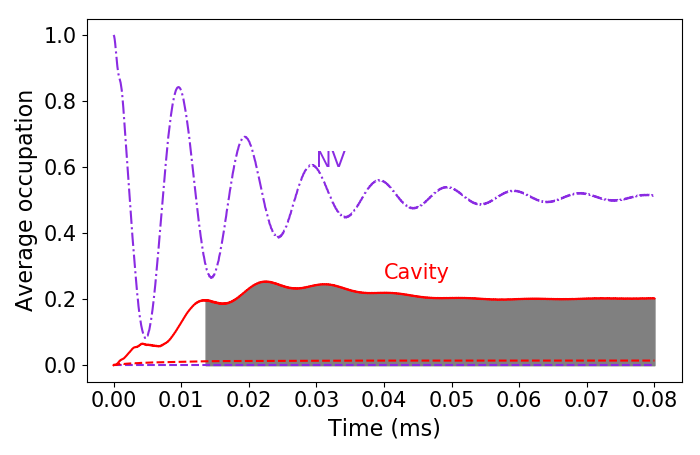}\\
\caption{(a) Periodic driving pulse scheme. The driving pulse is applied once the NV spin occupation is nearly 0, meaning that the spin state decays from $\ket{B}$ to $\ket{D}$. (b) Continuous driving pulse scheme. The NV spin and the cavity mode will reach a non-zero equilibrium state. The red solid curve and the purple dot-dashed curve represent cavity photon number and NV bright state ($\ket{B}$) population respectively. The gray shaded area corresponds to the time window for the readout operation, and the hatched area in (a) corresponds to the detection time window in one pulse cycle. The parameters used in (a) and (b) are the same as the ones in the entanglement generation section.} 
\label{fig:two_schemes}
\end{figure}

In the periodic driving scheme, periodic pulses are used to drive a cycling transition between the states $\ket{B}$ and $\ket{D}$. Assuming a perfect MW $\pi$ pulse is applied to the state $\ket{D}$, it is excited to the state $\ket{B}$ and then returns to the state $\ket{D}$ with a single photon emitted. Then we repeat this process. In the adiabatic elimination regime, the total Hamiltonian is given by
\begin{equation}
\hat{H}_{1} = \hat{H}_{\text{eff}} + g_{d}[\hat{\sigma}_{+} \exp(-i\omega_q t) + \hat{\sigma}_{-}\exp(i\omega_q t)]f(t),    
\end{equation} 
where $\hat{H}_{\text{eff}}$ is given by Eq.~(\ref{eq:adeha}), and $g_{d}$ is the coupling strength for the driving pulse, and $f(t)$ is a periodic delta function with the form $\delta(t - nT_{p})$ and the period $T_{p}$ is the inverse of the decay rate $R$. The simulation results are shown in Fig.~\ref{fig:two_schemes} (a). The solid red and dot-dashed purple curves are the cavity photon population and the NV spin population respectively when the NV spin is initially in the state $\ket{D}$, and the dashed red and purple lines are the cases where the initial NV spin state is $\ket{0}$. We can define the brightness (intensity) as the average number of emitted photons: $\beta_{i}=\kappa\int_{t_{0}}^{t_{0} + T}dt\langle \hat{a}^{\dagger}(t)\hat{a}(t) \rangle_i$ with $i = D$ or 0 representing the initial NV spin states in $\ket{D}$ and $\ket{0}$ respectively, where $\langle \hat{a}^{\dagger}(t)\hat{a}(t) \rangle_i$ is the corresponding average cavity photon number. A single photon is emitted within a period shown as the gray shade in Fig.~\ref{fig:two_schemes}(a).

To estimate the readout fidelity, we consider the measurement being repeated $N$ times and each measurement is independent. Thus, the number of photons detected within the total measurement time $NT_{p}$ can be described by a binomial distribution, and the probability of detecting $n$ photons is $P_{N, n, p} = {n\choose N} p^{n}(1 - p)^{N - n}$, where $p_{i} = \eta\beta_{i}$ is the probability of detecting a single photon within the detection time window, and $\eta$ is the total efficiency with which an emitted photon can be detected. One can plot $P_{N, n, p}$ corresponding to $\beta_D$ and $\beta_0$ and find the intersection point \cite{NoteX}. The intersection point is the threshold that decides the measurement result: if the number of photons detected is more than the threshold, the photons are most likely coming from the emitter and therefore the NV spin state is decided to be $\ket{D}$; if the number of photons detected are less than the threshold, the NV state is assumed to be $\ket{0}$ because these photons are highly possible from the thermal noise. The detailed discussion is in the supplementary material~\cite{NoteX}.

\subsection{Readout scheme using continuous driving pulses}
The continuous driving scheme employs a continuous-wave (CW) laser to drive the bright and the dark spin states. Similarly, the Hamiltonian in this case is given by 
\begin{equation}
\hat{H}_{1} = \hat{H}_{\text{eff}} + g_{d}(\hat{\sigma}_{+} \exp(-i\omega_q t)+\hat{\sigma}_{-}\exp(i\omega_q t)).  
\end{equation} 
Under this Hamiltonian, the cavity mode will eventually reach a non-zero equilibrium state as shown in Fig.~\ref{fig:two_schemes}(b). To give the calculation of the readout fidelity, we assume that the detection is a Poisson process, where the probability of detecting $n$ photons is given by  $P(n, \lambda) = \lambda^{n} e^{-\lambda}/n!$, where $\lambda$ is the average photon counts within total detection time $T_{0}$, given by $\lambda_{i}= \eta\kappa\int_{t_{0}}^{t_{0} + T_{0}}dt\langle \hat{a}^{\dagger}(t)\hat{a}(t) \rangle_i$ with $i = D$ or 0 corresponding to the initial states $\ket{D}$ or $\ket{0}$ respectively. Similarly to the treatment in the periodic driving scheme, the intersection point of these two plots of the probability distribution functions gives the threshold and the detailed discussion can be found in the supplementary material~\cite{NoteX}. 

\begin{figure}
\centering
\includegraphics[scale=0.3]{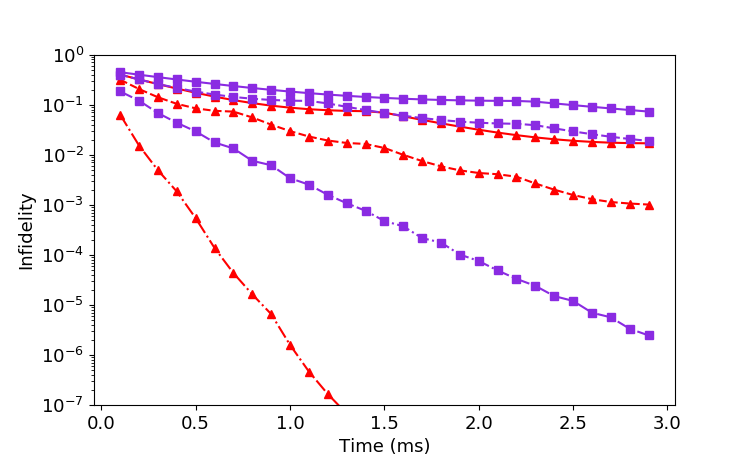}
\caption{The relation between the readout infidelity ($1-F$) and the total readout time with the parameters used in Fig.\ref{fig:two_schemes}. For the periodic driving scheme (plotted as purple squares), $\beta_{D} = 0.929$, $\beta_{0} = 0.034$, and the driving period (plotted as red triangles) is $T=0.02$ ms; for the continuous driving scheme, $\langle a^{\dag}(t)a(t) \rangle_D = 0.202$ and $\langle a^{\dag}(t)a(t) \rangle_0 = 0.014$. The solid, dashed, and the dash-dotted lines correspond to the total detection efficiency $\eta=$ 0.05, 0.1, and 0.5, respectively~\cite{Bai2017,doi:10.1063/1.3610677,yan2012ultra}. The time axis is the total readout time $NT_{p}$, where $N$ is the total pulse number in the periodic driving scheme. The discontinuity of the first derivative shown on the curves is due to the change of the threshold (because the threshold is always an integer).}
\label{fig:read_fid}
\end{figure}

Instead of showing the readout fidelity, here we show the readout infidelity ($1-F$) of these two schemes in Fig.~\ref{fig:read_fid} for the clearer demonstration of how well our readout schemes work. The dark count rate is taken to be 10~Hz in detectors~\cite{Bai2017}, which is negligible because the average number of dark counts within ms time period is on the order of $10^{-3}$, much smaller than the average number of emitted photons during the whole readout process. Also, the afterpulsing probability can be efficiently suppressed to be lower than 1$\%$~\cite{Bai2017}, which makes it negligible as well. Comparing these two schemes, the continuous driving scheme requires more time to have the same infidelity due to the lower signal-to-noise ratio in the present parameter regime than the periodic driving scheme. To achieve the high-fidelity readout ($>$ 99\%), the readout time is typically in the ms timescale for both of our schemes with detectors that have pretty poor efficiencies. However, a high-fidelity readout can be achieved in a shorter timescale if we use higher-efficiency detectors, which are however challenging to realize for telecom wavelength photons~\cite{doi:10.1063/1.3610677,doi:10.1063/1.2724816} at non-cryogenic temperatures. In comparison to other proposed methods~\cite{Jiang267,Shields2015,Liu2017,PhysRevB.100.125436,Siyushev2019,PhysRevX.9.021019}, which also demonstrate a high-fidelity readout of the electron spin in NV centers in ms timescale, these two readout schemes appear to predict comparable performance, without having to add extra elements to our setup. Thus, in our proposal for building a room-temperature quantum network, these spin-optomechanics system-based readout schemes serve as more natural and friendly candidates than other room-temperature readout methods.

\section{Repeater rates and overall fidelities}
\label{sec:rates}

We use a \enquote{two-round} repeater protocol. During the first round, the entanglement is generated between electron spins in every other elementary link and then is mapped to corresponding nuclear spins, which also sets those electron spins free. For the remaining links, the entanglement is generated in the second round, followed by the entanglement swapping that distributes entanglement between the first and last nuclear spins. Although entanglement generation between the electron spins is probabilistic, the failure of such an attempt does not disturb the entanglement stored in the nuclear spins if the dynamical decoupling is being applied during the entanglement generation~\cite{Fuchs2011,PhysRevLett.105.200402,PhysRevB.83.081201,Kalb2018}. This means that the second round of the entanglement generation process can be repeated many times until success. During the second round, the nuclear spin decoherence could degrade the stored entanglement from the first round, which is taken into consideration when computing the overall fidelities in Eq. (\ref{eq:overallfid}). Hence, our two-round repeater protocol makes the widely-used nested repeater structure no longer necessary~\cite{KimiaeeAsadi2018quantumrepeaters,Kumar_2019,RevModPhys.83.33}.

Considering an even number of links $m$, each with length $L_{0}$, the total entanglement distribution time is given by
\begin{equation}
\langle T \rangle_L=2f(m/2)\frac{L} {cmp_0} + T_{\text{mp}} + T_{\text{sw}},
\label{eq: nonmuxrates}
\end{equation}
where $f(m/2)$ is the factor of the average number of attempts required to successfully establish entanglement in all $m/2$ links, $p_0$ is the entanglement generation probability and $L$ is the total distance, and $c = 2\times10^{8}$ ms$^{-1}$ is the speed of light in optical fiber, and $T_{\text{mp}}$, $T_{\text{sw}}$ are the total entanglement mapping time and the total entanglement swapping time respectively. Both of these times are made up of $\mbox{C}\mbox{NOT}$ gate time plus the measurement time as discussed in Sec.~\ref{subsec: IIB} and Sec.~\ref{subsec:entswap}. The numerical results shown in the supplementary material~\cite{NoteX} show that $f(x) = 0.64\log_2(x) + 0.83$ is a good approximation, and one can recover the well-known $3/2$ factor by setting $x = 2$. In contrast to the nested repeater approach ~\cite{KimiaeeAsadi2018quantumrepeaters}, where the average entanglement distribution time has a linear dependence on the number of links, we here have a logarithmic dependence. Intuitively, the scaling improvement of the two-round protocol comes from the fact that there is no hierarchy of the entanglement swapping process, where higher-level swapping can only start under the condition of the success of the lower level. Therefore, the main thing left for us is to successfully generate the entanglement simultaneously for these links, which is calculated to have logarithmic dependence on $m/2$. This scheme could significantly enhance the entanglement distribution rate for a quantum network with much more links, e.g., networked quantum computing \cite{Nickerson2014}. Eq. (\ref{eq: nonmuxrates}) holds for nonmultiplexed repeaters, and the distribution time can be reduced using multiplexing. The multiplexing enables us to operate many parallel channels independently, and depending on which channel is finished first, we use the corresponding repeater. Thus, the probability of successfully establishing at least one channel is given by $1-(1-p_t)^N$ where $N$ is the number of multiplexing channels, and $p_t=mp_0/(2f(m/2))$ is the probability of establishing one channel. Therefore, the total entanglement distribution time is given by:

\begin{equation}
\langle T_{\text{mux}} \rangle_L=\frac{L/c}{1-(1-\frac{mp_0}{2f(m/2)})^N} + T_{\text{mp}} + T_{\text{sw}},
\label{eq: muxrates}    
\end{equation}
where $N$ is the number of multiplexing channels. When $N=1$, we have $\langle T_{\text{mux}} \rangle_L=\langle T \rangle_L$, which bring us back to Eq. (\ref{eq: nonmuxrates}). The multiplexing can be implemented either spatially or spectrally. For spatial multiplexing, we envision having many spin-optomechanical setups in each node \cite{PhysRevLett.98.060502}. The spectral multiplexing also requires many hybrid memories in one node but the emitted photons need to be converted to different frequencies fed into a common channel \cite{KimiaeeAsadi2018quantumrepeaters,Glorieux:12}. This can be accomplished using frequency translation which can be noise-free using waveguide electro-optic modulators \cite{PhysRevLett.119.083601}. The feeding to a common channel can be achieved by a tunable ring resonator filter that enables MHz-level resonance linewidths \cite{Yang2018}. 

\begin{figure}
    \centering
    $\hspace{-70mm}\mathbf{(a)}$\\
    \includegraphics[scale=0.55]{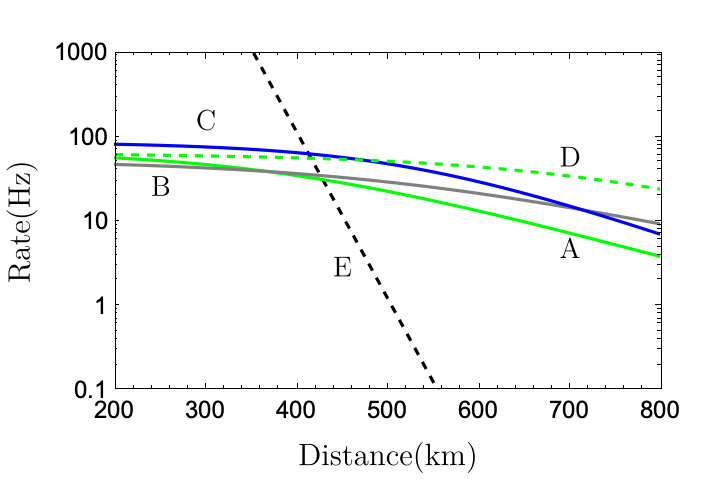}\\
    $\hspace{-70mm}\mathbf{(b)}$\\
    \includegraphics[scale=0.55]{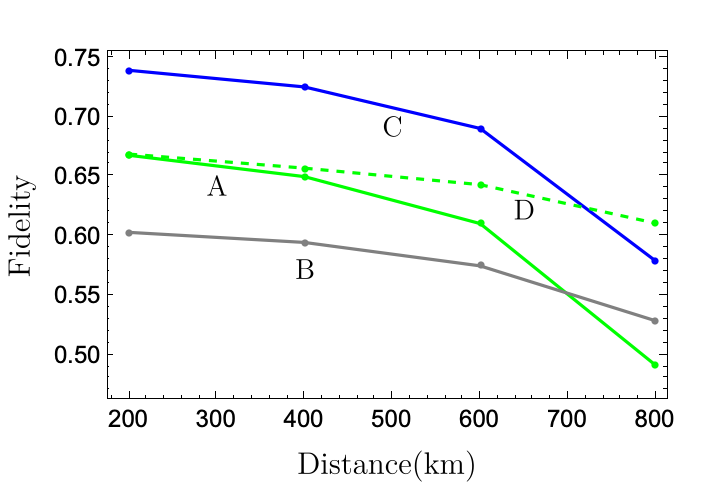}\\
    \caption{(a) Rates plots for the four optimal repeaters and direct transmission (black dashed line). $p_{0}=0.5p^{2}\eta_{d}^{2}\eta_{t}^{2}$ is the success probability of entanglement generation for the Barrett-Kok scheme with the success probability of emitting a single photon $p=0.9$, the transmission loss $\eta_t = \exp{(-L_{0}/2L_{\text{att}})}$ where $L_{0} = L/m$ is the length of each elementary link, and the detector efficiency $\eta_{d}=0.45$ ~\cite{Bai2017}. A is the 8-link repeater with $N=10$, B is the 10-link repeater with $N=10$, and C and D stand for the 6-link and 8-link repeaters with $N=100$ respectively, and $N$ stands for the number of multiplexing channels. The repetition rate of the single-photon source used in direct transmission is 10 GHz. (b) Fidelity plots for the four optimal repeaters with respect to the total distance with a detection efficiency of 45$\%$. The $\mbox{C}\mbox{NOT}$ gate fidelity is taken to be 99.2$\%$~\cite{Rong2015}. The electron spin readout fidelity is taken to be 99.9\% based on Fig. \ref{fig:read_fid}.}
    \label{Figure 9}
\end{figure}

Fig.~\ref{Figure 9}(a) shows the repeater rates as a function of distance for four optimal multiplexed repeaters and direct transmission. The repetition rate of the single-photon source direct transmission is assumed to be 10 GHz. These repeaters are the optimal choices from a set of repeaters investigated in the Supplemental material \cite{NoteX}. A is the 8-link repeater with $N=10$, B is the 10-link repeater with $N=10$, and C and D stand for the 6-link and 8-link repeaters with $N=100$ respectively. With $45\%$ detection efficiency, all these repeaters yield rates far above 1 Hz at 800 km. These rates are comparable to cryogenic schemes, such as the rare-earth ion-based scheme \cite{KimiaeeAsadi2018quantumrepeaters} and the microwave cat qubit-based scheme \cite{Kumar_2019}, and it outperforms the well-known DLCZ protocol for laser-cooling based systems \cite{RevModPhys.83.33}, which gives rates far below 1 Hz at 800 km with the same level of multiplexing, but the corresponding fidelities of the DLCZ protocol are higher as discussed later in this section. Even with similar fidelities, the rates of the DLCZ protocol are much lower than the rates of our protocol as the former is nested. However, if the detection efficiency is significantly lower, e.g. 10$\%$ \cite{yan2012ultra}, more multiplexing channels would be needed.

The whole repeater protocol consists of three parts described in Sec.~\ref{sec:quantum repeater}. However, instead of taking the fidelity of each part into consideration, here we consider the overall fidelity as
\begin{equation}
F_{\text{tot}}= \left(F_{\text{gen}}\right)^m \times \left(F_{\text{mp}}\right)^{m}\times \left(F_{\text{nro}}\right)^{m-1}\times e^{-\gamma_nt}, 
\label{eq:overallfid}
\end{equation}
where $F_{\text{gen}}$ is the fidelity of entanglement generation given in Fig.~\ref{Figure 4}, which needs to be established over $m$ elementary links. Note that $F_{\text{gen}}$ also depends on distance as the mechanically-induced thermal noise in the cavity effectively acts like dark counts. $F_{\text{mp}}$ is the fidelity of an entanglement mapping operation as described in Sec.~\ref{subsec: IIB}, and $F_{\text{nro}}$ is the readout fidelity of the nuclear spin. $\gamma_n$ is the decoherence rate of the nuclear spin, which is around 1 Hz, and $t$ is the total entanglement distribution time, i.e., either Eq. (\ref{eq: nonmuxrates}) for nonmultiplexed repeaters or Eq. (\ref{eq: muxrates}) for multiplexed repeaters. Then, the exponential factor $e^{-\gamma_n t}$ approximately characterizes the effect induced by the nuclear spin decoherence on the overall entanglement fidelity. Note that this overall fidelity equation is only valid in the high-fidelity regime where for each elementary link, the infidelities of entanglement generation ($1-F_{\text{gen}}$), entanglement mapping ($1-F_{\text{mp}}$), and the nuclear spin readout ($1-F_{\text{nro}}$) are much smaller than 1. This is the case here. The fidelity of entanglement swapping includes the fidelity of entanglement mapping plus the readout of two nuclear spins. Therefore, in total, we need to generate entanglement for $m$ links and perform $m$ times entanglement mapping to obtain a chain of nuclear spins followed by the readout of $m-1$ nuclear spins to achieve the final entangled state between the first and the last nuclear spins. The nuclear spin readout can be achieved by mapping its state to the electron spin and applying the readout methods discussed in Sec.~\ref{sec: readout}.

Fig.~\ref{Figure 9}(b) shows the overall fidelities with respect to the total distance for four different multiplexed repeaters with a detection efficiency of 45$\%$, the rates of which are shown in Fig.~\ref{Figure 9}(a). Again, these four repeaters are the optimal choices for yielding the best fidelities among the set of repeaters shown in Supplemental material \cite{NoteX}. As can be seen, increasing the number of multiplexing channels also increases the overall fidelities, which is attributed to the fact that the nuclear spin decoherence effect is alleviated when higher rates are achieved with a larger $N$. When $N=10$, the 8-link repeater A outperforms the 10-link repeater B from the crossover points (around 440 km shown Fig.~\ref{Figure 9}(a)) to 700 km but B takes over from 700 km to 800 km with a fidelity of around $53\%$ at 800 km. When $N=100$, the 6-link repeater C outperforms the 8-link repeater D from the crossover points (around 420 km) to 730 km but D takes over from 730 km to 800 km with a fidelity of around $61\%$ at 800 km. These fidelities are comparable to the fidelities of microwave cat qubit-based approach with around 60$\%$ for eight links \cite{Kumar_2019} but in general, they are lower than the fidelities of the DLCZ protocol for laser-cooling-based systems with 75$\%$ for eight links \cite{RevModPhys.83.33}, and cryogenic schemes such as the rare-earth ion-based scheme with around 80$\%$ for eight links. The overall entanglement fidelity could be further improved using entanglement purification protocols~\cite{PhysRevLett.76.722,Pan2001,Kalb928}, which would make this quantum network architecture fault-tolerant. It is worth noting that without multiplexing, the fidelities of the repeaters in this work are below $50\%$ (as shown in the Supplemental material), which cannot be further boosted using entanglement purification.

\section{Implementation}\label{sec:implementation}
The spin-optomechanics setup proposed in Ref.~\cite{Ghobadi2019} is mainly composed of a high-Q cavity patterning with a SiN membrane of ultrahigh Qf (quality$\times$frequency) product, where a small magnetic tip is attached. This hybrid device allows a single NV electron spin to be effectively coupled to photons inside the cavity, emitting a single photon with high purity and indistinguishability at room temperature. However, due to the design where the SiN membrane serves as a part of the optical cavity, the cavity finesse is limited to the order of $10^4$. The other key requirement for this system to work well is the low decay rate, $\kappa\sim 10^4~\text{Hz}$ in the optical cavity. These two key factors constrain the length of the cavity to be around 0.6m~\cite{Ghobadi2019}. Here, we propose a new design for this spin-optomechanics interface that uses the membrane-in-the-middle geometry to greatly reduce the cavity length. With this membrane-in-the-middle design, one could significantly reduce the cavity length using a high-finesse cavity, since the finesse scales as $F=\pi c/L\kappa$, where $\kappa$ is the cavity damping rate. As previously estimated, the cavity length is around $L=60$~\textrm{cm} with finesse $F=12000$. With the new design it might be possible to reduce this to around $L=0.6$~\textrm{cm}, if a finesse of order $10^6$ can be achieved, see e.g. Ref.~\cite{PhysRevA.64.033804}.

The spin-optomechanics interface shown in Fig.~\ref{Figure 1}(b) illustrates our envisioned spin-optomechanical transducer. A SiN membrane is placed between the node and the anti-node of the cavity modes (of both the cooling mode and the control mode) such that the optomechanical coupling is still linear and not quadratic like many other membrane-in-the-middle experiments~\cite{Thompson2008,Sankey2010,RevModPhys.86.1391}. The membrane-in-the-middle design allows us to use a membrane with a thickness much smaller than the light wavelength, which reduces the potential optical losses such as absorption and scattering due to the significantly smaller overlap between the membrane and the optical field~\cite{Thompson2008}. Similar to the previous proposal, a red-detuned control laser is used to drive the cavity for single photon extraction, which is set to be equal to the transition energy between dressed spin states $\omega_{\textrm{q}}$. The other red-detuned laser with detuning equal to the phonon sideband $\omega_\textrm{m}$ is used to cool the oscillator from room temperature, which is also possible to achieve in this proposed device~\cite{PhysRevLett.116.147202,Usami2012}. 

Moreover, the spin-mechanics coupling is achieved by a magnetic tip that is attached to the SiN membrane at the bottom, and a NV center in bulk diamond is placed nearby as shown in Fig.~\ref{Figure 1}(a). The required strong spin-mechanics coupling ($\lambda\sim10^5$ Hz) can be realized by a magnetic field gradient of $10^7$ T/m with a SiN membrane of $\sim$pg effective mass~\cite{Ghobadi2019}. This SiN membrane also needs to have ultra-low damping rate $\gamma_m$, which is discussed in~\cite{Ghadimi764,Ghobadi2019}. As the magnetic tip is attached to the SiN membrane, the quality factor of the membrane may be degraded. This could be compensated by further improving the initial quality factor of the membrane without the tip, which is possible to implement as the limit of the quality factor still has been not reached. With the combination of the methods in \cite{Ghadimi764} and \cite{PhysRevLett.116.147202}, one can get quality factors as high as $10^{10}$, which gives some room to improve our current Q factor $\sim 10^9$.

\section{Conclusions and outlook}\label{sec:conclusion}
We presented a room-temperature quantum network architecture based on NV centers in diamond and a spin-optomechanical interface. We showed that high-fidelity entanglement between electron spins can be generated between two distant nodes under realistic conditions. Nuclear spins associated with the NV centers can be utilized as quantum memories. We showed that the spin-optomechanical interface also offers the possibility to read out electron spins at room temperature with high fidelity on ms timescales. Furthermore, we proposed an entanglement distribution protocol where the average distribution time shows logarithmic scaling with the number of links as opposed to linear scaling in conventional nested protocols, and we discussed how multiplexing is an essential part of the protocol and showed how it can be used for boosting rates, allowing for feasible final fidelities at long distances. A membrane-in-the-middle design may allow to reduce the dimensions of the spin-optomechanics interface to the sub-cm range, thus improving its potential for integration and scalability.

We have here focused on room-temperature quantum repeaters as a medium-term goal, but the proposed approach also holds promise for the implementation of distributed quantum computing \cite{Nickerson2013,Nickerson2014}, extending photonic approaches to quantum information processing in diamond ~\cite{PhysRevX.4.031022,PhysRevX.5.031009} beyond cryogenic temperatures. Nuclear spins in diamond offer the possibility to implement quantum error correction codes~\cite{Kalb928,Taminiau2014,PhysRevLett.96.070504,PhysRevLett.112.250501}, which, when integrated into our present approach, may enable fault-tolerant quantum communication and quantum computation under ambient conditions. 


\section*{Acknowledgements}
\label{sec:acknowledgements}
We thank Sumit Goswami for the helpful and stimulating discussions, and we thank Evan Meyer-Scott for pointing out the mistake in the attenuation length, which led to this revised version. This work was supported by the Natural Sciences and Engineering Research Council of Canada (NSERC) through its Discovery Grant (DG), Canadian Graduate Scholarships (CGS), CREATE, and Strategic Project Grant (SPG) programs; and by Alberta Innovates Technology Futures (AITF) Graduate Student Scholarship (GSS) program. S.C.W. also acknowledges support from the SPIE Education Scholarship program. 



\section*{Supplemental material}\label{sup}
\subsection*{S1: System Hamiltonian and dissipation}
The spin-optomechanics system has the following Hamiltonian ($\hbar=1$):
\begin{equation}
\begin{aligned}
\hat{H}&=\omega_q \hat{\sigma}_+\hat{\sigma}_-+\omega_q \hat{a}^{\dagger}\hat{a}+\omega_m \hat{b}^{\dagger}\hat{b}+\lambda(\hat{b}^{\dagger}\hat{\sigma}_- +\hat{b} \hat{\sigma}_+)\\
&+g(\hat{b}^{\dagger}\hat{a} +\hat{b} \hat{a}^{\dagger}),
\end{aligned}
\end{equation}
where we already performed rotating wave approximations and the cooling laser mode $\hat{c}$ is ignored as it just cools the mechanical oscillator to be close to the ground state, converting phonons to photons that are emitted at a different frequency than the desired single photon from the NV spin. In order to perform the adiabatic elimination, we need to convert the Hamiltonian into the natural picture~\cite{brion2007adiabatic} by entering the rotating frame using the following transformation:
\begin{equation}
\hat{H}_1=\hat{U}\hat{H}\hat{U}^{\dagger}-i\hat{U}\Dot{\hat{U}}^{\dagger}, \end{equation}
where $\hat{U}=e^{i\omega_q(\hat{a}^{\dagger}\hat{a}+\hat{\sigma}_+\hat{\sigma}_-+\hat{b}^{\dagger}\hat{b})t}$. Then, one obtains the Hamiltonian
\begin{equation}
\hat{H}_1=\delta \hat{b}^{\dagger}\hat{b}+\lambda(\hat{b}^{\dagger}\hat{\sigma}_- +\hat{b} \hat{\sigma}_+)+g(\hat{b}^{\dagger}\hat{a} +\hat{b} \hat{a}^{\dagger}),  
\label{Hamil}
\end{equation}
where $\delta=\omega_m-\omega_q$ is the detuning between the control laser (and the dressed NV spin) and the phonon sideband. Taking dissipation into consideration, the master equation is given by
\begin{equation}
\begin{aligned}
\Dot{\hat{\rho}}=&-i[\hat{H}_1,\hat{\rho}]+\kappa\mathcal{D}[\hat{a}]\hat{\rho}+\gamma^*_s\mathcal{D}[\hat{\sigma}_z]\hat{\rho}+n_{\text{th}}\gamma_m\mathcal{D}[\hat{b}^{\dagger}]\hat{\rho}\\
&+(n_{\text{th}}+1)\gamma_m\mathcal{D}[\hat{b}]\hat{\rho},
\end{aligned}
\label{master}
\end{equation}
where the intrinsic NV spin flip-flop rate is ignored because it is much smaller than the spin dephasing rate $\gamma^*_s$ in an isotopically purified diamond~\cite{Balasubramanian2009}.

Then, the corresponding Heisenberg-Langevin equations are given by
\begin{equation}
\begin{aligned}
\label{equ:hes_lan}
&\Dot{\hat{a}}=-\frac{\kappa}{2} \hat{a} -ig \hat{b} \\
&\Dot{\hat{b}}=-i\delta \hat{b}- \frac{\gamma_m}{2}  \hat{b} -ig \hat{a} -i\lambda\hat{\sigma}_-  + \sqrt{\gamma_m}\hat{F}_{b}(t)\\
&\Dot{\hat{\sigma}}_-=-2\gamma^*_s \hat{\sigma}_- +i\lambda \hat{\sigma}_z \hat{b},
\end{aligned}
\end{equation}
where $\hat{F}_{b}(t)$ is the noise operator that satisfies
\begin{equation}
    \begin{split}
    \label{equ:noise_corr}
        \langle \hat{F}_{b}^{\dagger}(t)\hat{F}_{b}(t^{\prime}) \rangle = n_{\text{th}}\delta(t - t^{\prime}).
    \end{split}
\end{equation}

\vspace{1mm}
\subsection*{S2: Adiabatic elimination}
When $\delta\gg \lambda,g$, one can adiabatically eliminate the oscillator either by following the method \cite{zhang2015proposal} to obtain the Heisenberg-Langevin equations for cavity mode $\hat{a}$ and NV spin $\hat{\sigma}_-$ after the elimination of $\hat{b}$ or by setting $\Dot{\hat{b}}=0$, and obtaining $\hat{b}$ in terms of $\hat{a}$ and $\hat{\sigma}_-$. Here, we follow the second way to obtain
\begin{equation}
\hat{b}=\frac{ig\hat{a}+i\lambda \hat{\sigma}_--\sqrt{\gamma_m}\hat{F}_b(t)}{-i\delta-\gamma_m/2}.    
\end{equation}
Under the conditions $\delta\gg\gamma_m/2$ and $\gamma_m\ll 1$, which are true in this system, this can be well approximated as
\begin{equation}
\hat{b}\approx\frac{g\hat{a}+\lambda \hat{\sigma}_-}{-\delta},
\label{effmode}
\end{equation}
where we ignore decay-related terms and only keep coherent parts. Now, substituting this in the Hamiltonian (Eq. (\ref{Hamil})), we obtain the effective Hamiltonian after the adiabatic elimination
\begin{equation}
\label{equ:eff_h}
\hat{H}_{\text{eff}}=\frac{\lambda^2}{\delta}\hat{\sigma}_+\hat{\sigma}_-+\frac{g^2}{\delta}\hat{a}^{\dagger}\hat{a}+\Omega(\hat{a}^{\dagger}\hat{\sigma}_-+\hat{a}\hat{\sigma}_+),    
\end{equation}    
where $\Omega=\lambda g/\delta$ is the effective interaction between the cavity mode and the NV electron spin. In order to get the effective master equation, we also need to compute the decoherence terms related to the oscillator mode $\hat{b}$. Using Eq.~(\ref{effmode}), the thermal relaxation Lindbladian $(n_{\text{th}}+1)\gamma_m\mathcal{D}[\hat{b}]\hat{\rho}$ can be rewritten as 
\begin{widetext}
\begin{equation}
\begin{aligned}
&(n_{\text{th}}+1)\gamma_m\Big[(\frac{g\hat{a}+\lambda \hat{\sigma}_-}{-\delta})\hat{\rho}(\frac{g\hat{a}^\dagger+\lambda \hat{\sigma}_+}{-\delta})-(\frac{g\hat{a}^\dagger+\lambda \hat{\sigma}_+}{-\delta})(\frac{g\hat{a}+\lambda \hat{\sigma}_-}{-\delta})\hat{\rho}/2-\hat{\rho}(\frac{g\hat{a}^\dagger+\lambda \hat{\sigma}_+}{-\delta})(\frac{g\hat{a}+\lambda \hat{\sigma}_-}{-\delta})/2\Big]\\
\approx &(n_{\text{th}}+1)\gamma_m\Big[\frac{g^2}{\delta^2}\hat{a}\hat{\rho}\hat{a}^\dagger+\frac{\lambda^2}{\delta^2}\hat{\sigma}_-\hat{\rho}\hat{\sigma}_+-\frac{g^2}{\delta^2}\hat{a}^\dagger\hat{a}\hat{\rho}/2-\frac{\lambda^2}{\delta^2}\hat{\sigma}_+\hat{\sigma}_-\hat{\rho}/2-\frac{g^2}{\delta^2}\hat{\rho}\hat{a}^\dagger\hat{a}/2-\frac{\lambda^2}{\delta^2}\hat{\rho}\hat{\sigma}_+\hat{\sigma}_-/2\Big]\\
=&\frac{g^2}{\delta^2}(n_{\text{th}}+1)\gamma_m\mathcal{D}[\hat{a}]\hat{\rho}+\frac{\lambda^2}{\delta^2}(n_{\text{th}}+1)\gamma_m\mathcal{D}[\hat{\sigma}_-]\hat{\rho},
\end{aligned}
\end{equation}
\end{widetext}
where the off-diagonal terms correspond to the incoherent interaction between the cavity mode and the spin and the thermal-induced cross-decoherence between these two modes, which can be ignored if $\delta\gg n_{\text{th}}\gamma_m$. This is satisfied in our system even at ambient conditions. The same is true for the thermal excitation Lindbladian $n_{\text{th}}\gamma_m\mathcal{D}[\hat{b}^\dagger]\hat{\rho}$, which can be written as
\begin{equation}
n_{\text{th}}\gamma_m\mathcal{D}[\hat{b}]\hat{\rho}\approx \frac{g^2}{\delta^2}n_{\text{th}}\gamma_m\mathcal{D}[\hat{a}^\dagger]\hat{\rho}+\frac{\lambda^2}{\delta^2}n_{\text{th}}\gamma_m\mathcal{D}[\hat{\sigma}_+]\hat{\rho}.
\end{equation}

Therefore, the effective master equation is given by
\begin{equation}
\begin{aligned}
\Dot{\hat{\rho}}=&-i[\hat{H}_{\text{eff}},\hat{\rho}]+\kappa_1\mathcal{D}[\hat{a}]\hat{\rho}+\kappa_2\mathcal{D}[\hat{a}^{\dagger}]\hat{\rho}+\gamma^*_s\mathcal{D}[\hat{\sigma}_z]\hat{\rho}\\
&+\gamma_1\mathcal{D}[\hat{\sigma}_-]\hat{\rho}+\gamma_2\mathcal{D}[\hat{\sigma}_+]\hat{\rho},
\end{aligned}
\label{effmaster}
\end{equation}
where $\kappa_1=\kappa+g^{2}\gamma_m(n_{\text{th}}+1)/\delta^2$ is the effective cavity decay rate, and  $\kappa_2=g^{2}n_{\text{th}}\gamma_m/\delta^2$, $\gamma_1=\lambda^2\gamma_m(n_{\text{th}}+1)/\delta^2$, and  $\gamma_2=\lambda^2n_{\text{th}}\gamma_m/\delta^2$ are the mechanically-induced thermal excitation rate for the cavity mode, and the mechanically-induced thermal flip-flop rates for the spin respectively.

\subsection*{S3: Effective emission rate}
Under the condition $\lambda = g$, the effective Hamiltonian shown in Eq.~(\ref{equ:eff_h}) can be rewritten in the rotating frame of the spin frequency $\lambda^2/\delta$
\begin{equation}
    \hat{H}_{\text{int}} = \Omega(\hat{a}^{\dagger}\hat{\sigma}_{-} + \hat{a}\hat{\sigma}_{+}).
\end{equation}
Together with the effective master equation shown in Eq.~(\ref{effmaster}), we obtain a set of optical Bloch equations for the cavity photon population, NV spin population, and the coherence between them as
\begin{widetext}
\begin{equation}
    \begin{split}
        \frac{d\langle\hat{a}^{\dagger}\hat{a}\rangle}{dt} & = i\Omega(\langle \hat{a}\hat{\sigma}_{+}\rangle - \langle  \hat{a}^{\dagger}\hat{\sigma}_-\rangle) - (\kappa_{1} - \kappa_{2}) \langle \hat{a}^{\dagger}\hat{a}\rangle +\kappa_2,\\
        \frac{d\langle \hat{\sigma}_{+}\hat{\sigma}_{-} \rangle}{dt} & = i\Omega(\langle \hat{a}^{\dagger}\hat{\sigma}_-\rangle-\langle \hat{a}\hat{\sigma}_{+}\rangle) - (\gamma_{1} +\gamma_{2})\langle \hat{\sigma}_{+} \hat{\sigma}_{-} \rangle+\gamma_2,\\
        \frac{d\langle \hat{a}^{\dagger}\hat{\sigma}_- \rangle}{dt} & = i\Omega(\langle \hat{a}^{\dagger} \hat{a}\hat{\sigma}_z\rangle + \langle \hat{\sigma}_+ \hat{\sigma}_-\rangle) - \frac{\kappa_{1} - \kappa_{2}}{2}\langle \hat{a}^{\dagger}\hat{\sigma}_{-}\rangle
        - \frac{
        \gamma_{1} + \gamma_{2}}{2} \langle \hat{a}^{\dagger}\hat{\sigma}_{-}\rangle - 2\gamma^*_{s}\langle \hat{a}^{\dagger}\hat{\sigma}_{-}\rangle,\\
        \frac{d\langle \hat{a}\hat{\sigma}_{+} \rangle}{dt} & = -i\Omega(\langle \hat{a}^{\dagger} \hat{a}\hat{\sigma}_z\rangle +\langle \hat{\sigma}_+ \hat{\sigma}_-\rangle) - \frac{\kappa_{1} - \kappa_{2}}{2}\langle \hat{a}\hat{\sigma}_{+}\rangle
        - \frac{
        \gamma_{1} + \gamma_{2}}{2} \langle \hat{a}\hat{\sigma}_{+}\rangle - 2\gamma^*_{s}\langle \hat{a}\hat{\sigma}_{+}\rangle. \\
    \end{split}
\end{equation}
Since we are mainly interested in the single-photon regime, the term $\langle \hat{a}^{\dagger} \hat{a}\hat{\sigma}_z\rangle$ can be simplified as $-\langle \hat{a}^{\dagger}\hat{a}\rangle$. Hence, these optical Bloch equations can be rewritten as
\begin{equation}
    \begin{split}
        \frac{d\langle\hat{a}^{\dagger}\hat{a}\rangle}{dt} & = i\Omega(\langle \hat{a}\hat{\sigma}_{+}\rangle - \langle  \hat{a}^{\dagger}\hat{\sigma}_-\rangle) - (\kappa_{1} - \kappa_{2}) \langle \hat{a}^{\dagger}\hat{a}\rangle +\kappa_2,\\
        \frac{d\langle \hat{\sigma}_{+}\hat{\sigma}_{-} \rangle}{dt} & = i\Omega(\langle \hat{a}^{\dagger}\hat{\sigma}_-\rangle-\langle \hat{a}\hat{\sigma}_{+}\rangle) - (\gamma_{1} +\gamma_{2})\langle \hat{\sigma}_{+} \hat{\sigma}_{-} \rangle+\gamma_2,\\
        \frac{d\langle \hat{a}^{\dagger}\hat{\sigma}_- \rangle}{dt} & = i\Omega(\langle \hat{\sigma}_+ \hat{\sigma}_-\rangle-\langle \hat{a}^{\dagger}\hat{a}\rangle) - \frac{\kappa_{1} - \kappa_{2}}{2}\langle \hat{a}^{\dagger}\hat{\sigma}_{-}\rangle - \frac{
        \gamma_{1} + \gamma_{2}}{2} \langle \hat{a}^{\dagger}\hat{\sigma}_{-}\rangle - 2\gamma^*_{s}\langle \hat{a}^{\dagger}\hat{\sigma}_{-}\rangle,\\
        \frac{d\langle \hat{a}\hat{\sigma}_{+} \rangle}{dt} & = -i\Omega(\langle \hat{\sigma}_+ \hat{\sigma}_-\rangle-\langle \hat{a}^{\dagger}\hat{a}\rangle) - \frac{\kappa_{1} - \kappa_{2}}{2}\langle \hat{a}\hat{\sigma}_{+}\rangle - \frac{
        \gamma_{1} + \gamma_{2}}{2} \langle \hat{a}\hat{\sigma}_{+}\rangle - 2\gamma^*_{s}\langle \hat{a}\hat{\sigma}_{+}\rangle. \\
    \end{split}
\end{equation}
\end{widetext}

In the incoherent regime, the cross terms that are responsible for the Rabi oscillation, i.e., $\langle \hat{a}^{\dagger}\hat{\sigma}_-\rangle$ and $\langle \hat{a}\hat{\sigma}_+\rangle$, can be eliminated \cite{auffeves2010controlling}, resulting in 
\begin{equation}
    \begin{split}
        \frac{d\langle \hat{a}^{\dagger}\hat{a}\rangle}{dt} & = -(R + \kappa_{1} - \kappa_{2})\langle \hat{a}^{\dagger}\hat{a} \rangle + R\langle \hat{\sigma}_{+}\hat{\sigma}_{-} \rangle +\kappa_2,\\
        \frac{d\langle \hat{\sigma}_{+}\hat{\sigma}_{-} \rangle}{dt} & = -(R + \gamma_{1}+\gamma_{2})\langle \hat{\sigma}_{+}\hat{\sigma}_{-} \rangle + R\langle \hat{a}^{\dagger}\hat{a}\rangle +\gamma_2,
    \end{split}
\end{equation}
where R is the effective decay rate which describes the population transfer between the cavity photon and the NV spin, and it is given by
\begin{equation}
    R = \frac{4\Omega^{2}}{\kappa_{1} - \kappa_{2} + \gamma_{1} + \gamma_{2} + 2\gamma^*_{s}}.
\end{equation}
Moreover, given that at room temperature $n_{\text{th}}\gg 1$, the effective decay rate $R$ can be written in a more compact form
\begin{equation}
R= \frac{4\Omega^{2}}{\kappa + 2\gamma^*_{s} + 2\Gamma_{\text{th}}},   
\end{equation}
where $\Gamma_{\text{th}}=\lambda^2n_{\text{th}}\gamma_m/\delta^2=\lambda g n_{\text{th}}\gamma_m/\delta^2$ is the thermal noise for the NV electron spin.

\subsection*{S4: Initial state of the cavity}
The initial state can be obtained by solving the steady state of cavity mode with only the optomechanical coupling $g$ turned on. Thus, we set $\Omega=0$, and we obtain the following equation:
\begin{equation}
\frac{d\langle\hat{a}^{\dagger}\hat{a}\rangle}{dt} = - (\kappa_{1} - \kappa_{2}) \langle \hat{a}^{\dagger}\hat{a}\rangle +\kappa_2=0.    
\end{equation}
Solving this equation, we get the average occupation number of the cavity mode: $\bar{n}_c=\langle \hat{a}^{\dagger}\hat{a}\rangle=\frac{\kappa_2}{\kappa_1-\kappa_2}$. As this occupation is very small $\bar{n}_c\approx 10^{-3}$, it is valid to truncate the Hilbert space up to $\ket{1}$. Hence, the initial state of the cavity is given by:
\begin{equation}
\rho_{ic}= \frac{\kappa_1-2\kappa_2}{\kappa_1-\kappa_2}\ket{0}\bra{0}+\frac{\kappa_2}{\kappa_1-\kappa_2}\ket{1}\bra{1}.   
\end{equation}

\subsection*{S5: Photon counting statistics}
\label{appendix:readout_stats}

Our goal is to distinguish spin states $\ket{D}$ and $\ket{0}$. Let us denote the conditional probabilities of measurement outcome $\pm$ given that the initial state of system is $\ket{i}$, with $i\in\{D,0\}$, as $P(\pm|i)=p_i^\pm$. The total probability of outcome $\pm$ is then given by $p^\pm = p_D p_D^+ + p_0 p_0^+$ where $p_i$ is the total probability of the system being in state $i$. Then the conditional fidelity is defined as the conditional probability $P(D|+)$ ($P(0|-)$) of having state $D$ ($0$) given outcome $+$ ($-$). This is given by Bayes' theorem: $F^+\equiv P(D|+)=p_D p^+_D/p^+$ and $F^-\equiv P(0|-)=p_0 p_0^-/p^-$. We can then define the total fidelity as the weighted average $F=(p^+F^+ + p^-F^-)/p_\eta$ where $p_\eta = p^++p^-$ is the total probability of having a measurement outcome. In the case that $p_D=p_0=1/2$ and $p_\eta=1$, the fidelity reduces to the average of the conditional probabilities $F = (p^+_D+p^-_0)/2$.

\begin{figure}
    \centering
    \includegraphics[scale=0.43]{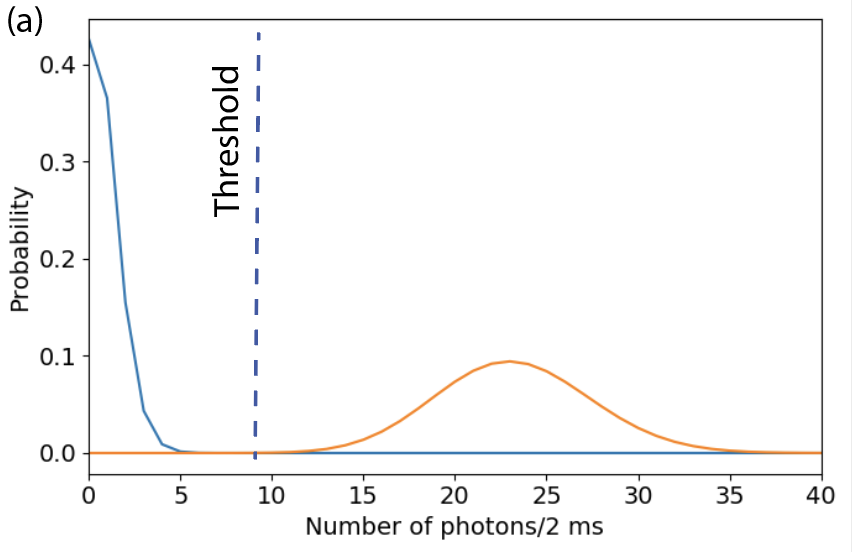}
    \includegraphics[scale=0.43]{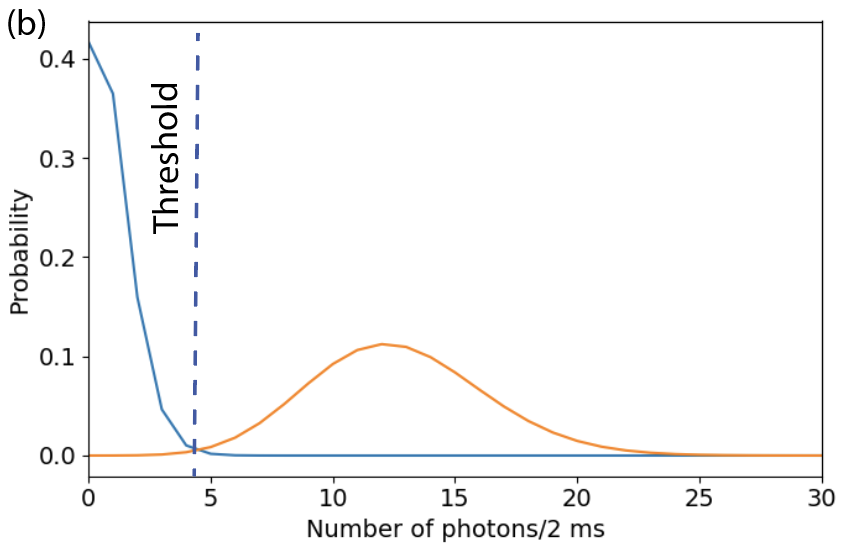}
    \caption{Photon-counting histogram for pulsed driving scheme (a) and the continuous driving scheme (b) with a total readout time of 2 ms. The threshold is determined by the intersection of the two plots.}
    \label{fig:pdh}
\end{figure}

The most widely-used approach for spin readout is to use a cycling transition, which involves the emission and detection of a large number of photons. The photon-counting histogram shows the probability distribution of the number of photons detected and has two traces: one for photons emitted from the emitter and the other for the thermal noise contribution (non-zero cavity photon number when the spin state is at $\ket{0}$). The cross-over point of the two traces corresponds to the photon number threshold, above which we can be confident that the photons come from the emitter, thus determining that the spin state is $\ket{D}$; otherwise, the spin state is $\ket{0}$, meaning that the photons most likely come from the thermal noise.

Here we show the photon-counting histogram for the pulsed and continuous driving schemes in Fig. \ref{fig:pdh}. For the pulsed driving scheme, the photon-counting histogram is described by a binomial distribution $P_{N, n, p} = {n\choose N} p^{n}(1 - p)^{N - n}$, where $p = \eta\beta$, $\eta$ is the total efficiency that an emitted photon can be detected, and $\beta$ is the brightness of the cavity photon. For the parameters used in Fig. \ref{fig:two_schemes}, $\beta = 0.929$ and $\beta=0.034$ for the initial spin states $\ket{D}$ and $\ket{0}$ respectively. We plot the photon-counting histogram in Fig. \ref{fig:pdh}(a) for a total pulse number of 100 (so the corresponding total readout time is 2 ms). The blue solid line and the yellow solid line show the probability distribution with respect to the detected photon number when the spin is in state $\ket{D}$ and $\ket{0}$, respectively. The threshold is thus determined by the corresponding number of photons at the intersection of the two lines, and it is $n_t = 9$ in this case. The readout fidelity is given by
\begin{equation}
    F = \frac{1}{2}(\sum_{n < n_{t}} P_{N, n, p_{2}} + \sum_{n \geq n_{t}} P_{N, n, p_{1}}).
    \label{fid_calculation}
\end{equation}
Then the estimated fidelity is $0.99999$.

For the continuous driving scheme, we plot the photon-counting histogram for the corresponding Poisson distribution, shown in Fig. \ref{fig:pdh}(b). In this case, the probability distribution of detecting $n$ photons is $P(n, \lambda) = \lambda^{n}e^{-\lambda}/n!$, where $\lambda$ is the average number of photons detected and is proportional to the readout time. For the parameters we used in Fig. \ref{fig:read_fid}, $\lambda_{D}/\lambda_{0} = 14.43$, where $\lambda_{D}$ and $\lambda_{0}$ are for the case of spin state $\ket{D}$ and $\ket{0}$, respectively. This gives two probability distributions that 
intersect at a photon number of 4. This means the threshold is 4, and the readout fidelity is $0.997$ using Eq. (\ref{fid_calculation}).

\subsection*{S6: $f(x)$ Derivation}
\label{sec:appendix_fx}
\begin{table}
\centering
    \caption{Numerical results of $f(x)$}
    \begin{tabular}{c | c | c | c | c | c }
    x & 2 & $2^{2}$ & $2^{3}$ & $2^{4}$ & $2^{5}$\\
    \hline
    $f(x)$ & 1.5 & 2.08 & 2.72 & 3.38 & 4.05\\
    \end{tabular}
    \label{tab:cons}
\end{table}

Here we provide a derivation of $f(x)$ used in Sec. \ref{sec:rates}. For $x$ elementary links, we define the average number of attempts required to independently generate entanglement in all $x$ links as $n_{max, x} = f(x)/p_{0}$, where $p_{0}$ is the entanglement generation probability. For a single link, the probability of a successful entanglement generation with $n$ attempts is given by $P(n) = p_{0}(1 - p_{0})^{n-1}$. Thus the joint probability of successful entanglement generation for all $x$ links with attempts $n_{1}, n_{2}, ..., n_x$ is
\begin{equation}
\begin{aligned}
P_j(n_{1}, n_{2}, ..., n_{x}) &= \prod_{k = 1}^{x} P(n_{k})\\
         &= p_{0}^{x}(1 - p_{0})^{\sum_{k = 1}^{x}n_{k} - x}.    
\end{aligned}
\end{equation}
The probability distribution function (PDF) of $n_{max, x}$ is
\begin{widetext}
\begin{equation}
\begin{split}
\label{equ:pdf_nmax}
    P(n_{max, x}) & = \sum_{k = 1}^{x} P_j(n_{k} = n_{max, x},n_{\neq k} < n_{max, x})
    + \sum_{k = 1,l=2}^{l,x} P_j(n_{k, l} = n_{max, x}, n_{\neq k \neq l} < n_{max, x}) \\
     & + ... +  P_j(n_{1} = n_{2} = ... = n_{x} = n_{max, x}).
\end{split}
\end{equation}
\end{widetext}

However, it is difficult to calculate $n_{max, x}$ from Eq.~(\ref{equ:pdf_nmax}). To simplify the problem, we assume $x = 2^{k}$. The PDF of $n_{max, x} = n_{max, 2^k}$ can be calculated iteratively by separating $2^k$ links into two groups of sublinks with each having $2^{k-1}$ sublinks. $n_{max1, 2^{k-1}}$ and $n_{max2, 2^{k-1}}$ denote the number of attempts for these two sublinks respectively. Then the probability distribution of $n_{max, x}$ can be expressed as

\begin{widetext}
\begin{equation}
\begin{aligned}
  P_j(n_{max1, 2^{k-1}}, n_{max2, 2^{k-1}}) &= P(n_{max1, 2^{k-1}}) * P(n_{max2, 2^{k-1}}),\\
P({n_{max, 2^{k}}}) &= P_{j}(n_{max1, 2^{k-1}} = n_{max, 2^{k }},
     n_{max2, 2^{k-1}} < n_{max, 2^{k }})\\
     &+ P_{j}(n_{max1, 2^{k-1}} < n_{max, 2^{k}},
    n_{max1, 2^{k}} = n_{max, 2^{k}}. 
\end{aligned}
\end{equation}

The simplest case is $k=1$,
\begin{equation}
\begin{aligned}
     P_{j}(n_1, n_2)  &= P(n_1) P(n_2), \\
     P(n_{max, 2})  &= P_{j}(n_{1} = n_{max, 2}, n_2 < n_{max, 2})\\
      &+ P_{j}(n_{1} < n_{max, 2}, n_2 = n_{max, 2}).\\
\end{aligned}
\end{equation}
\end{widetext}
We numerically calculate $f(n)$ with respect to $k = 1$ to $5$, shown in Tab.~\ref{tab:cons}.

One can check that the function $f(2^k)$ almost linearly increases with $k$, and the regression result gives 
\begin{equation}
    f(2^{k}) = 0.64k + 0.83.
\label{eq: numerics}    
\end{equation}
Therefore, we obtain the following empirical expression for $f(x)$ by replacing $2^k$ with $x$ and $k$ with $\log_{2}(x)$ in Eq. (\ref{eq: numerics}).
\begin{equation}
    f(x) = 0.64\log_{2}(x) + 0.83.
\end{equation}

\subsection*{S7: Optimal choice of repeater rates and fidelities}
\label{sec:appendix_rates}
\begin{figure*}
\centering
\includegraphics[scale=0.65]{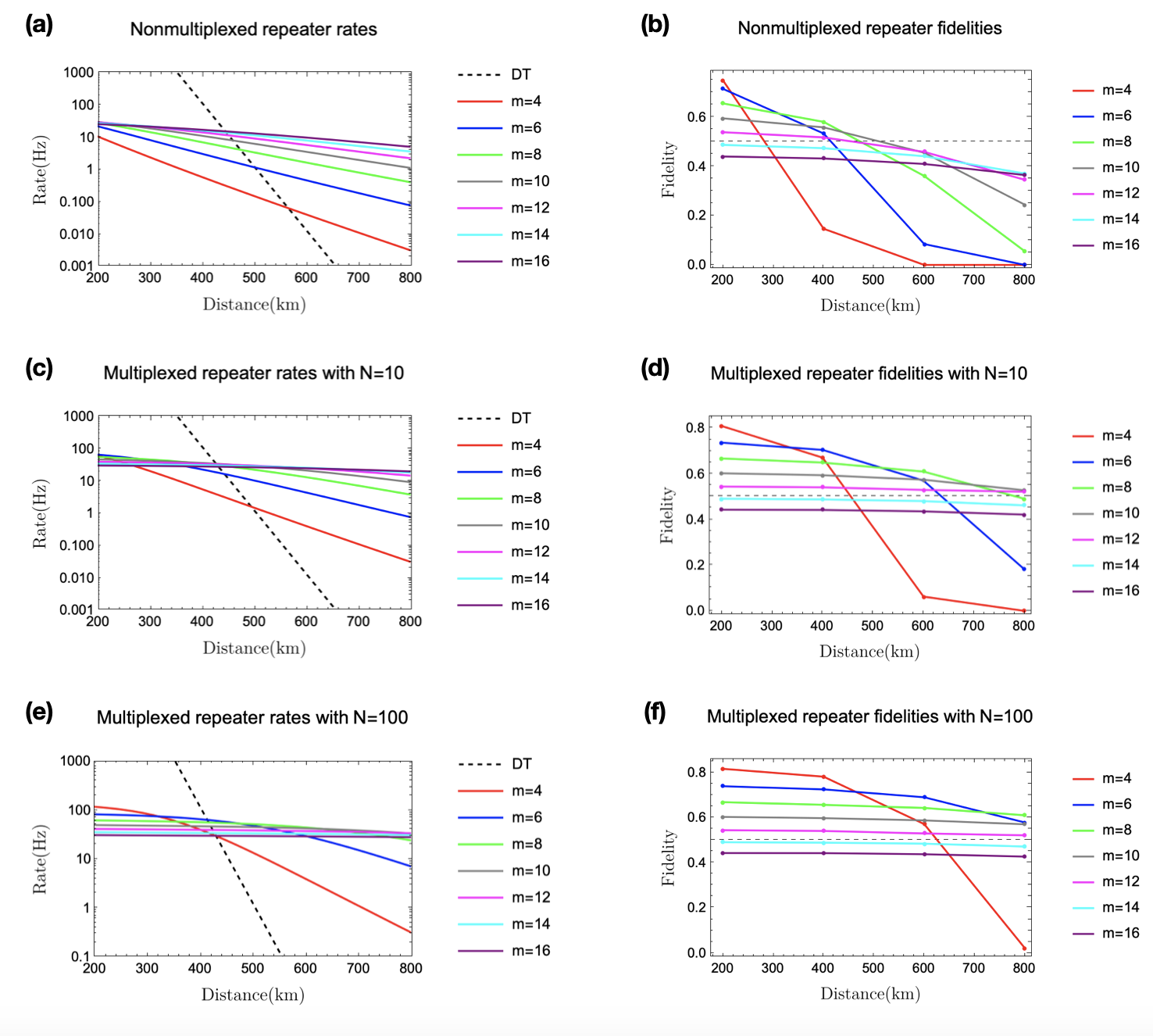}
\caption{(a), (c), and (e) are the rates as a function of total distance for nonmultiplexed repeaters, multiplexed repeaters with 10 channels, and multiplexed repeaters with 100 channels. $m$ is the number of links, and in total there are seven different numbers of links are shown. DT stands for direct transmission (black dashed line). (b), (d), and (f) are the fidelities as a function of total distance for nonmultiplexed repeaters, multiplexed repeaters with 10 channels, and multiplexed repeaters with 100 channels. The horizontal dashed line in (b), (d), and (f) stands for the $50\%$ purification threshold.}
\label{performance figs}    
\end{figure*}

Here we plot the repeater rates and fidelities versus the total distance for three different scenarios: no multiplexing, multiplexing by 10 channels ($N=10$), and multiplexing by 100 channels ($N=100$). In (a), the nonmultiplexed repeaters with different numbers of links outperform direct transmission (black dashed line) at different lengths, and the 16-link repeater yields the highest rates with a rate close to $10$ Hz when $L=800$ km. However, the nonmultiplexed repeaters have quite low fidelities as shown in (b) where almost all repeaters drop below $50\%$ after 500 km. The rates of the multiplexed repeaters shown in (c) are all higher than those in (a), and at $L=800$ km, the 12-link, 14-link, and 16-link repeaters yield rates above 10 Hz. For the same multiplexed repeaters in (d), the fidelities are enhanced compared to the fidelities in (b) due to the enhanced rates that alleviate the nuclear spin decoherence. When reaching 800 km, only the 10-link and 12-link repeaters have fidelities above $50\%$, and the former is the optimal choice if we desire to distribute entanglement to 800 km. However, the 8-link repeater is the optimal choice between the distance at which direct transmission is beaten and 700 km. (e) and (f) show the rates and fidelities of the multiplexed repeaters with 100 channels. For the rates, they are further increased due to more multiplexing channels, and even at $L=800$ km, only the 4-link repeater drops below 1 Hz, and other repeaters yield rates close to and above 10 Hz. For the fidelities of these multiplexed repeaters shown in (f), the 6-link, 8-link, 10-link, and 12-link repeaters achieve final fidelities above $50\%$ at 800 km with the 8-link repeater yielding a fidelity of around $60\%$, and the 8-link repeater is the optimal choice when the total distance lies between around 730 km and 800 km. However, if the total distance is shorter than 730 km, the 6-link repeater is the optimal choice.

\bibliographystyle{apsrev4-1}
\bibliography{mybib}

\end{document}